\documentclass[aps,prb,reprint,superscriptaddress]{revtex4-2}

\usepackage[english]{babel}
\usepackage[version=4]{mhchem}  
\usepackage{amsmath}  
\usepackage{lipsum}
\usepackage{amsfonts} 
\usepackage{graphicx}  
\graphicspath{{pic/}}  
\usepackage{varioref} 
\usepackage{braket}  
\usepackage{dsfont}  
\usepackage{siunitx}  
\usepackage{hyperref} 
\usepackage{xcolor}

\newcommand{\abs}[1]{\left|#1\right|}  

\begin{document}


\title{Propagation of longitudinal acoustic phonons in \ce{ZrTe5} exposed to a quantizing magnetic field}


\author{Toni Ehmcke}
\affiliation{Institute for Theoretical Physics and W\"urzburg-Dresden Cluster of Excellence ct.qmat, Technische Universit\"at Dresden, 01069 Dresden, Germany.}

\author{Stanislaw Galeski}
\affiliation{Max Planck Institute for Chemical Physics of Solids, 01187 Dresden, Germany.}

\author{Denis Gorbunov}
\affiliation{Hochfeld-Magnetlabor Dresden (HLD-EMFL) and W\"urzburg-Dresden Cluster of Excellence ct.qmat, Helmholtz-Zentrum Dresden-Rossendorf, 01328 Dresden, Germany}

\author{Sergei Zherlitsyn}
\affiliation{Hochfeld-Magnetlabor Dresden (HLD-EMFL) and W\"urzburg-Dresden Cluster of Excellence ct.qmat, Helmholtz-Zentrum Dresden-Rossendorf, 01328 Dresden, Germany}

\author{Joachim Wosnitza}
\affiliation{Hochfeld-Magnetlabor Dresden (HLD-EMFL) and W\"urzburg-Dresden Cluster of Excellence ct.qmat, Helmholtz-Zentrum Dresden-Rossendorf, 01328 Dresden, Germany}
\affiliation{Institut f\"ur Festk\"orper- und Materialphysik, Technische Universit\"at Dresden, 01062 Dresden, Germany}

\author{Johannes Gooth}
\affiliation{Max Planck Institute for Chemical Physics of Solids, 01187 Dresden, Germany.}
\affiliation{Institut f\"ur Festk\"orper- und Materialphysik, Technische Universit\"at Dresden, 01062 Dresden, Germany}

\author{Tobias Meng}
\affiliation{Institute for Theoretical Physics and Würzburg-Dresden Cluster of Excellence ct.qmat, Technische Universität Dresden, 01069 Dresden, Germany.}


\date{\today}

\begin{abstract}
The compound \ce{ZrTe5} has recently been connected to a charge-density-wave (CDW) state with intriguing transport properties. Here, we investigate quantum oscillations in ultrasound measurements that microscopically originate from electron-phonon coupling and analyze how these would be affected by the presence or absence of a CDW. We calculate the phonon self-energy due to electron-phonon coupling, and from there deduce the sound-velocity renormalization and sound at\-te\-nu\-a\-tion. We find that the theoretical predictions for a metallic Dirac model resemble the experimental data on a quantitative level for magnetic fields up to the quantum-limit regime.
\end{abstract}


\maketitle

\section{Introduction} \label{sec:intro}
The three-dimensional (3D) compound zirconium pentatelluride (\ce{ZrTe5}) has been extensively studied due to its large negative magnetoresistivity\cite{li_chiral_2016, chen_magnetoinfrared_spec_2015}, thermoelectric properties \cite{zhang_anomalous_2019, wang_first_2018, jones_thermoelectric_1982} and because it possibly exhibits a variety of topological phases, including weak and strong topological insulator and Dirac semimetal phases \cite{weng_transition-metal_2014, zhang_electronic_2017, xu_temperature_2018, liu_zeeman_2016, wu_contactless_2019}. A focus of recent discussions has been the observation of plateau-like features in the transverse resistivity accompanied by minima in the longitudinal resistivity both in \ce{ZrTe5} \cite{liang_anomalous_2018, tang_three_dimensional_2019, galeski_origin_2021} and the structurally similar \ce{HfTe5} \cite{galeski_unconventional_2020, wang_approaching_2020}. These plateau-like features have been proposed to be linked to the formation of a periodic modulation of electron density, a charge-density wave (CDW), and to a truly quantized Hall effect in three dimensions.

Alternatively, it has been proposed that the Hall response of \ce{ZrTe5} may only be quasi-quantized, and results from a gapless state without a CDW \cite{galeski_origin_2021}. It is thus worthwhile to search for additional experimental probes allowing to discriminate between a gapped CDW state and a quasi-quantized metallic state. As one key consequence, the formation of a CDW would induce a gap in the electronic spectrum, and, therefore, strongly suppress the electronic density of states at the Fermi level. In principle, this has experimentally observable consequences in transport \cite{okada_negative_1982, trescher_cdw_2017, trescher_quantum_2018}, magnetization  \cite{mahan_many_2010}, and phonon dynamics \cite{jericho_velocity_1980, song_instability_2017, saintpaul_elastic_2016}. Furthermore, a CDW amplitude mode can directly be probed by Raman spectroscopy \cite{tsang_raman_1976}. While some of the existing experimental data has been argued to agree with theoretical models that predict the existence of a CDW transition in \ce{ZrTe5} \cite{Qin_cdw_2020,Zhao.2021}, other experiments on $\ce{ZrTe5}$ show negative evidence for a CDW state \cite{okada_negative_1982, tian_gap_opening_2021, galeski_origin_2021}. To falsify one of the two scenarios, it is thus desirable to have access to a detailed modeling of a large number of complementary observables. In this work, we focus on a quantitative analysis of ultrasound measurements that we presented alongside in Ref.~\cite{galeski_origin_2021} to further strengthen the hypothesis of a quasi-quantized metallic state in \ce{ZrTe5}. We do so by including electron-phonon interactions to the very same model that has been used to describe the transport and magnetization measurements in Ref.~\cite{galeski_origin_2021}. We complement these findings with a calculation that adds a CDW into our model, and show that a CDW would only be consistent with the experimental data if the electron-phonon coupling was unexpectedly large.

\begin{figure}[h!]
	\centering 
	\includegraphics[width=\linewidth]{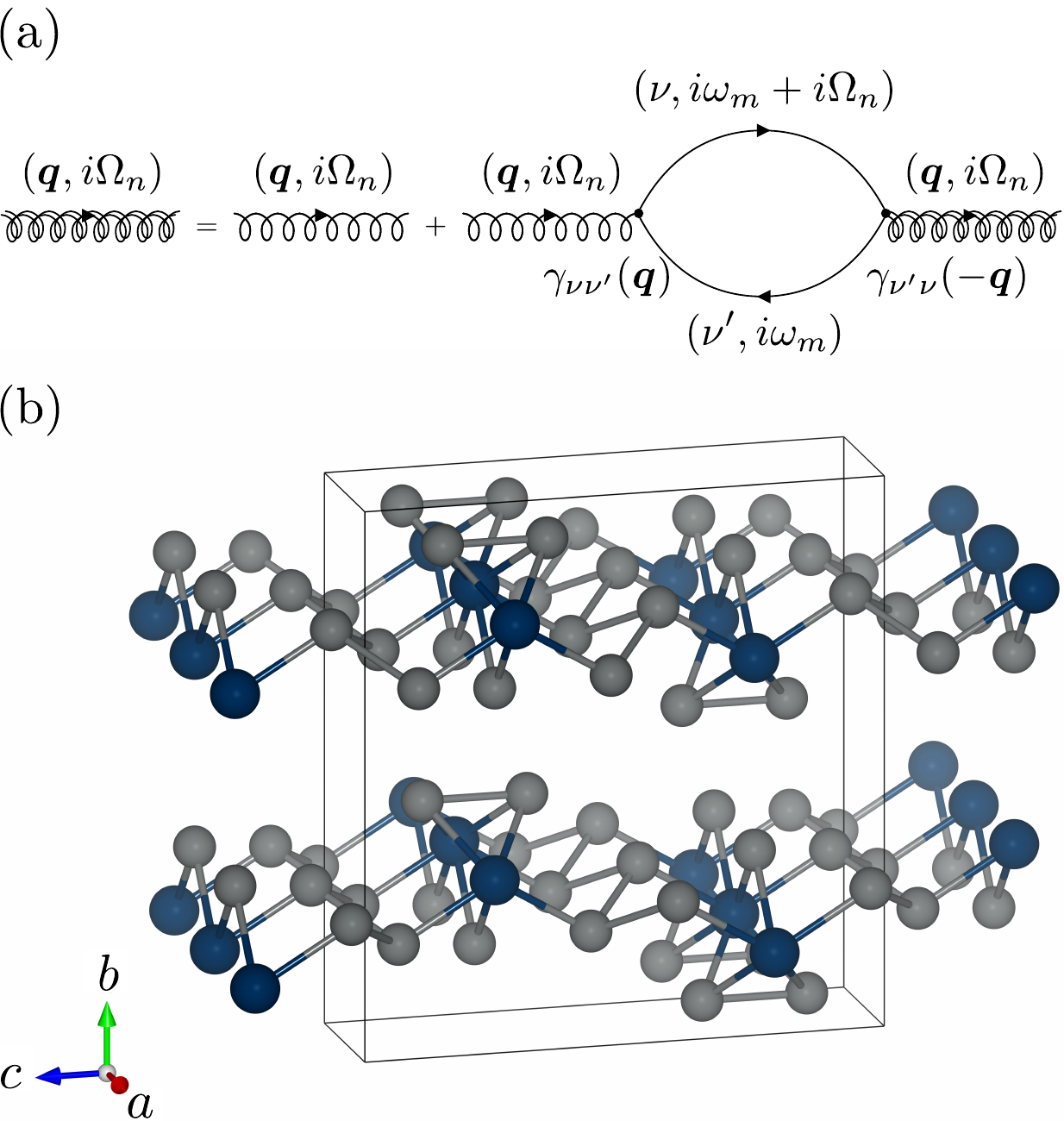} 
	\caption{(a) Perturbative expansion for the full propagator (double-curled) of the longitudinal acoustic phonon mode that couples to Dirac fermions in a magnetic field via the electron-phonon interaction given in Eq.~\eqref{eq:el_ph_int} in the spirit of a random-phase approximation. A free phonon (single-curled) may decay into a particle-hole pair of free Dirac fermions (solid) and recombines eventually to a new phonon. (b) Orthorhombic crystal structure of \ce{ZrTe5}. The \ce{Zr} atoms form trigonal prismatic chains of \ce{ZrTe3} that are connected via zigzig chains of \ce{Te} atoms (grey). These weakly coupled, quasi-two-dimensional sheets of \ce{ZrTe5} stack along the $b$ axis.} 
	\label{fig:feynman_lattice} 
\end{figure} 

We theoretically investigate longitudinal acoustic phonons in \ce{ZrTe5}, and, in particular, how their propagation is affected by external magnetic fields. Due to the electron-phonon interaction, lattice vibrations can decay into electron-hole pairs that eventually recombine to form a new phonon as shown in Fig.~\ref{fig:feynman_lattice}(a). As a consequence, the lattice vibrations inherit a magnetic-field dependence from the intermediate electron-hole-pair propagation. Thus, phonon-related observables, such as Hall viscosity \cite{shapourian_visco_2015}, Raman scattering \cite{rinkel_signatures_2017}, sound velocity \cite{zhang_quantum_2020, schindler_strong_2020, laliberte_field_2020, rinkel_influence_2019}, and sound attenuation \cite{spivak_magneto_2016, pikulin_chiral_2016, Antebi2021, Sukhachov2021}, are powerful tools for monitoring the electronic structure of a system. We focus our analysis on the renormalized sound velocity and the sound attenuation resulting from electron-phonon coupling in \ce{ZrTe5}.

The remainder of the paper is organized as follows. In Sec.~\ref{sec:framework}, we construct the effective low-energy Hamiltonian for the electronic degrees of freedom in \ce{ZrTe5} in the presence of a static, homogeneous magnetic field. Furthermore, we introduce phonons to the system, and we explain how these couple to the electrons. Using a random-phase-like approximation, we obtain the dressed phonon propagator. We then deduce the sound-velocity renormalization and the sound attenuation, which we evaluate and compare to experimental data in Sec.~\ref{sec:results}. Finally in Sec.~\ref{sec:cdw},  we compare the results of the metallic Dirac scenario with a model containing a CDW in the quantum-limit regime of the system. This allows us to analyze the effect of a bulk gap on the phonon self-energy.          

\section{Theoretical framework} \label{sec:framework}
\subsection{From microscopics to an effective low-energy Landau-band theory} \label{sec:landau_bands} 
The microscopic crystal structure of \ce{ZrTe5} is depicted in  Fig.~\ref{fig:feynman_lattice}(b). This compound exhibits the orthorhombic  space group \textit{Cmcm} (number 63) \cite{furuseth_crystal_1973} and the crystallographic lattice axes $(a, b, c)$ define the coordinate system $(x, z, y)$ that we use in the remainder. The material is layered along the $z$ axis. Overall, the electronic band structure exhibits the usual degree of complexity typically encountered in solids, with numerous bands deriving from multiple orbitals at each atom in the unit cell. At low energies, however, this complexity is reduced dramatically. Various theoretical and experimental studies have shown that the low-energy band structure of \ce{ZrTe5} exhibits a (possibly weakly gapped) Dirac node at the $\Gamma$ point \cite{tang_three_dimensional_2019, weng_transition-metal_2014}. In addition, several other pockets might be present in the Brillouin zone, but consensus has not yet been reached on their existence and physical consequences (they might, for example, be important for the temperature dependence of transport) \cite{zhang_electronic_2017, chenjie_thermodynamically_2021}. Sample-to-sample variations and growth conditions might play some role in explaining these apparent discrepancies.

Our study is geared towards the description of ultrasound experiments in a specific set of samples, namely the ones reported in Ref.~\cite{galeski_origin_2021}. The most important feature in this data for the present discussion are plateau-like structures in the Hall response of \ce{ZrTe5}, which have also been seen in independent transport measurements \cite{tang_three_dimensional_2019}, and which are well-captured by the model. Moreover, since the model includes all symmetry-allowed terms, we expect our findings to be generically important for understanding whether or not \ce{ZrTe5} has a CDW. In the considered samples, quantum oscillations show no signatures of Fermi-surface sheets in addition to the Dirac pocket around $\Gamma$. Furthermore, even the inclusion of quadratic bands at the boundary of the first Brillouin zone has been argued to not change the electronic-transport properties close to the quantum-limit qualitatively at low temperatures \cite{chenjie_thermodynamically_2021}. As we discuss below, also ultrasound measurements in these samples can be explained by only taking into account a single Dirac pocket. The possible presence of additional Fermi surfaces, thus, plays no significant role for transport and ultrasound properties in our case. 

These observations motivate a model that focusses on the low-energy electrons in the Dirac pocket close to the $\Gamma$ point. Given the space group and symmetries of the material, the most general Bloch Hamiltonian up to first order in the three-dimensional momentum $\mathbf{k}$ has been shown to read \cite{chen_magnetoinfrared_spec_2015}
\begin{align} \label{eq:h_zero_field}
		h_{e}(\boldsymbol{k}) = m\, \tau_3 \sigma_0 + \hbar (	&v_x k_x\, \tau_1 \sigma_3\nonumber \\
																&+ v_y k_y\, \tau_2 \sigma_0 + v_z k_z\, \tau_1 \sigma_1)\,,
\end{align}
where $\tau_i$ ($\sigma_i$) are Pauli matrices describing an orbital (spin) degree of freedom, $m$ is a mass parameter, $v_i$ are Fermi velocities, and $\hbar$ is the reduced Planck constant. Given the layered character of the material, it is not surprising that the Dirac pocket is strongly elongated (cigar-like) along the $k_z$ axis, which means that $v_x, v_y \gg v_z$.

In the remainder, we are in particular interested in the effects of a static, homogeneous magnetic field $\boldsymbol{B}=B\boldsymbol{e}_z$ along the $z$ direction on the ultrasound properties (we choose $B>0$). While phonons do not couple directly to the magnetic field, the electron-phonon interaction mediates an indirect coupling of the phonons with the magnetic field passing via the electrons. For the latter, the effect of a magnetic field is modeled by including the Zeeman term $h_Z = -g\mu_BB\,\tau_0\sigma_3 / 2$ and the orbital magnetic effect $\hbar\boldsymbol{k}\rightarrow\hbar\boldsymbol{k}+e\boldsymbol{A}$ \cite{chen_magnetoinfrared_spec_2015}. Here, $g$ is the \textit{g}-factor, $\mu_B$ is the Bohr magneton, $-e$ is the electronic charge and $\boldsymbol{A}=-yB\boldsymbol{e}_x$ is the vector potential in Landau gauge. The second-quantized electronic Hamiltonian, therefore, reads
\begin{align}\label{eq:h_finite_field_sq}
	\mathcal{H}_e&=H_e-\mu N=\int\mathrm{d}^3 r\, \psi(\boldsymbol{r})^\dagger [h_e(\boldsymbol{r})-\mu]\,\psi(\boldsymbol{r})\,,\\
	\label{eq:h_finite_field}
	h_e(\boldsymbol{\hat{r}})&= m \tau_3 \sigma_0 + v_x \left(\hbar\hat{k}_x - eB\hat{y}\right)\, \tau_1 \sigma_3 \nonumber\\
		&+ \hbar\left(v_y \hat{k}_y\, \tau_2 \sigma_0 + v_z \hat{k}_z\, \tau_1 \sigma_1\right) 
		-\frac12 g\mu_BB\,\tau_0\sigma_3\,,
\end{align}
where $\hat{k}_i = -i\hat\partial_{x_i}$, $\mu$ is the chemical potential, $\psi(\boldsymbol{r})=(\psi_{+\uparrow}(\boldsymbol{r}), \psi_{+\downarrow}(\boldsymbol{r}), \psi_{-\uparrow}(\boldsymbol{r}), \psi_{-\downarrow}(\boldsymbol{r}))^T$ is the vector of annihilation operators for electrons with orbital quantum number $\pm$ and spin quantum number $\uparrow,\downarrow$ (defined with respect to~$\tau_3$ and $\sigma_3$, respectively), and $N = \int\mathrm{d}^3 r\, \psi(\boldsymbol{r})^\dagger \psi(\boldsymbol{r})$ denotes the particle number.

Diagonalizing the electronic Hamiltonian yields the {Landau band structure} of the system. Each Landau band has the typical degeneracy of $L_x L_y / (2\pi\ell_B^2)$,  where $L_x$ ($L_y$) is the length of the system in $x$ direction ($y$ direction), and where $\ell_B = \sqrt{\hbar/(eB)}$ is the magnetic length.  For $v_x < 0$ and $v_y > 0$, the {zeroth} Landau bands have the dispersion
\begin{equation}\label{eq:lbs_zero}
	\epsilon_{0s}(k_z)=s\sqrt{m_*^2 + (\hbar v_z k_z)^2}\,.
\end{equation}
with $s=\pm$, and where the renormalized Dirac gap is $m_*=m - \frac12 g \mu_B B$. The Zeeman term is thus the only coupling between the magnetic field and the zeroth Landau bands.  The {higher} Landau bands, on the contrary, have energies
\begin{align}\label{eq:lbs}
		\epsilon_{nst}(k_z)=s\left[\vphantom{\frac12}\right.	&\left(\sqrt{m^2+2e\hbar B|v_x|v_y n}+\frac12 tg \mu_B B\right)^2\nonumber \\
								 	&+ (\hbar v_z k_z)^2\left.\vphantom{\frac12}\right]^\frac12\,,
\end{align}
with $n = 1, 2, \dots\,,$ and $s,t=\pm$. The calculation of the electron-phonon coupling in Sec.~\ref{sec:el_ph_int} also requires the eigenspinors of the Hamiltonian \eqref{eq:h_finite_field}, which can be found in Appendix \ref{app:eigenstates}.

Diagonalizing the many-body Hamiltonian \eqref{eq:h_finite_field_sq}, we obtain
\begin{align}
	\mathcal{H}_e&=\sum_{k_xk_zs}
	\begin{aligned}[t]
		\left[\vphantom{\sum_t}\right.	&\xi_{0s}(k_z)\,c_{k_xk_z0s}^\dagger c_{k_xk_z0s}\\
																	&+\sum_{nt} \xi_{nst}(k_z)\,c_{k_xk_znst}^\dagger c_{k_xk_znst}\left.\vphantom{\sum_t}\right]
	\end{aligned}\\
	&\equiv \sum_\nu \xi_\nu c_\nu^\dagger c_\nu\,,
\end{align}
where $\nu$ is a combined index of all electronic quantum numbers that classifies the fermionic eigenmodes $c_\nu$ and eigenenergies $\xi_\nu = \epsilon_\nu - \mu$. 
The propagation of the Dirac electrons is described by their Matsubara Green's functions
\begin{equation}\label{eq:free_el_gf}
	\mathcal{G}^{(0)}(\nu;\tau)=-\braket{\hat{\mathcal{T}}_\tau c_\nu(\tau)c_\nu(0)^\dagger}_0\,.
\end{equation}
Here, $\hat{\mathcal{T}}_\tau$ is the imaginary-time-ordering operator, $\braket{\dots}_0=\operatorname{Tr}(e^{-\beta \mathcal{H}_e}\dots)/\operatorname{Tr}(e^{-\beta \mathcal{H}_e})$ is the thermal average with respect to $\mathcal{H}_e$ at inverse thermal energy $\beta=1/(k_BT)$ and $O(\tau)=e^{\tau \mathcal{H}_e/\hbar}O e^{-\tau \mathcal{H}_e/\hbar}$ is the Heisenberg-picture time evolution for an operator $O$. Following Endo et al. \cite{Endo_2009}, we modify the bare Green's functions ad-hoc by a finite self-energy that represents disorder and impurity scattering in the system. We thus use the Green's functions
\begin{equation}\label{eq:el_gf_imp}
	\mathcal{G}^{(0)}(\nu;i\omega_n)\rightarrow\frac{1}{i\hbar \omega_n - \xi_\nu - \Sigma_\nu(i\omega_n)}\,,
\end{equation}
where $\Sigma_\nu(i\omega_n)$ is the electronic self-energy contribution due to impurity scattering and $\omega_n = (2n + 1)\pi / (\hbar \beta)$ are fermionic Matsubara frequencies. Although a microscopic calculation of the full impurity-averaged electronic Green's function from Eq.~\eqref{eq:el_gf_imp} would yield a non-trivial frequency dependence \cite{Klier_2015, koenye_magnetoresistance_2018}, we replace the self-energy by the constant Landau-level broadening, $\Sigma_\nu(i\omega_n) = -i\operatorname{sgn}(\omega_n)\,\Gamma$. This ad-hoc approximation surely restricts the regime of validity for the following considerations to the limit of weak impurity scattering and magnetic fields up to the quantum-limit regime, i.e., for fields $B\leq \SI{1.5}{T}$ \cite{tang_three_dimensional_2019, galeski_origin_2021}. However, it allows for an easy interpretation of results and successfully described observables previously for various quantum-Hall-type systems \cite{galeski_origin_2021, Endo_2009, hsu_anomalous_2010} or in the context of Weyl electrons \cite{ashby_chiral_2014} and paramagnetic states in pyrochlore iridates \cite{rhim_quantum_2015}. In the following, we choose either $\Gamma = \SI{0.1}{\milli\electronvolt}$ (see Sec.~\ref{sec:cdw}) or $\Gamma=\SI{0.5}{\milli\electronvolt}$ (see Sec.~\ref{sec:results}).

\subsection{Fixing the charge-carrier density}
Having defined our effective low-energy model, we are now in a position to discuss the density of the charge carriers in the system. Since we assume the samples to be electrically isolated,  the number of charge carriers is fixed, also as a function of the magnetic field. Because the valence band of our linearized low-energy model is formally occupied by an infinite number of electrons, we define the particle number with respect to the filled Dirac sea (chemical potential $\mu$ at the Dirac node, $\mu = 0$).  The electronic density relative to this reference state is
\begin{align}
	n_0 &=\frac{1}{V}\sum_{\substack{\nu \\ \epsilon_\nu > 0}} \braket{c_\nu^\dagger c_\nu}-\frac{1}{V}\sum_{\substack{\nu \\ \epsilon_\nu < 0}} \braket{c_\nu c_\nu^\dagger}\\
	&= \frac{1}{V}\int_{-\infty}^{\infty}\mathrm{d}\omega\hbar \sum_{\nu} \operatorname{sgn}(\epsilon_\nu)\rho_\nu(\omega)n_F(\operatorname{sgn}(\epsilon_\nu)\hbar\omega)\label{eq:part_dens}\,,
\end{align}
where $n_F(\epsilon)=\left[\exp(\beta \epsilon) + 1\right]^{-1}$ is the Fermi function, and the spectral function $\rho_\nu(\omega) = -\operatorname{Im}G_r^{(0)}(\nu;\omega)/\pi$ is defined from the retarded Green function $G_r^{(0)}(\nu;\omega)=\mathcal{G}^{(0)}(\nu; i\omega_n\rightarrow\omega + i0^+)$. In the zero-temperature limit, where $n_F(\epsilon)=\Theta(-\epsilon)$, the frequency integration in Eq.~\eqref{eq:part_dens} is performed analytically,
\begin{widetext}
	\begin{align}
			n_0 = \frac{1}{2\pi\ell_B^2}\int_{-\infty}^{\infty}\frac{\mathrm{d}k_z}{2\pi}
				\left\{\vphantom{\frac12}\right.	
					&\arctan\left(\frac{\epsilon_{0+}(k_z)+\mu}{\Gamma}\right) -\arctan\left(\frac{\epsilon_{0+}(k_z)-\mu}{\Gamma}\right)\nonumber\\
					&+\sum_{nt}\left[\arctan\left(\frac{\epsilon_{n+t}(k_z)+\mu}{\Gamma}\right) -\arctan\left(\frac{\epsilon_{n+t}(k_z)-\mu}{\Gamma}\right)\right]
				\left.\vphantom{\frac12}\right\}\,.\label{eq:part_dens_t0}
	\end{align}
\end{widetext}

It turns out that this expression diverges with respect to the $n$ series after the ad-hoc introduction of a spectral broadening $\Gamma > 0$. Because the band structure of \ce{ZrTe5} is no longer described by a Dirac Hamiltonian at energies far above the Fermi level, however, our model naturally requires a high-energy cut-off $\Lambda$ that regularizes Eq.~\eqref{eq:part_dens_t0}. This also fixes the problem of a divergent particle number. Crucially, we find that the ultrasound properties analyzed in the present work are essentially cut-off independent (see Appendix \ref{app:cut-off}).  In the remainder, we choose $\Lambda=\SI{25}{\milli\electronvolt}$, which is sufficiently large compared to the chemical potential in the maximal considered range of magnetic-field strengths $B\leq\SI{10}{\tesla}$ and certainly sufficient in the range $B\leq\SI{3}{\tesla}$, for which we compare theory and experiments.

\begin{figure}
	\centering
	\includegraphics[width=\columnwidth]{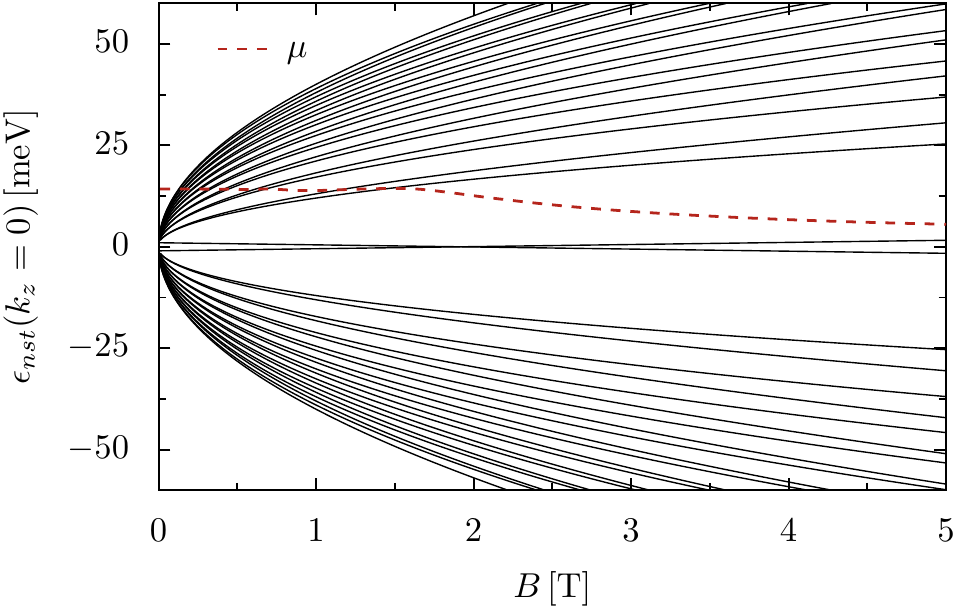}
	\caption{Landau-band extrema (black, solid) $\epsilon_{0s}(0)$ and $\epsilon_{nst}(0)$ from Eqs.~\eqref{eq:lbs_zero} and \eqref{eq:lbs}, respectively, as a function of magnetic-field strength and the chemical potential (red, dash-dotted) that fixes the charge-carrier density from Eq.~\eqref{eq:part_dens}. At small fields, the Fermi level is approximately fixed. Fluctuations in $\mu$ become significant for $B\geq\SI{1.5}{\tesla}$, the quantum limit regime. The parameters are given by: $m=\SI{1}{\milli\eV},\,v_z=\SI{15340}{\m\per\s},\,g=18,\,\sqrt{2e\hbar|v_x|v_y}=\SI{12.5}{\milli\eV\tesla\tothe{-\frac12}},\,n_0=\SI{2e-7}{\per\cubic\angstrom},\,T=\SI{2}{\kelvin},\,\Gamma=\SI{0.5}{\milli\electronvolt},\,\Lambda=\SI{25}{\milli\electronvolt}\,.$}
	\label{fig:lbs}
\end{figure}

In order to pin the charge-carrier density, the Fermi energy has to shift as a function of the magnetic field, $\mu=\mu(B)$. The corresponding self-consistently calculated chemical potential is shown in Fig.~\ref{fig:lbs} as a function of the magnetic field along with the minima of the Landau bands. We find that the Fermi energy only shows small fluctuations around the value at zero field. This is due to an interplay between the field dependence of the Landau bands and their increasing degeneracy. The quantum-limit regime in which only the zeroth Landau band is occupied is reached at about $B_{\text{QL}}=\SI{1.5}{\tesla}$, which is remarkably small compared to other materials.

\subsection{Electron-phonon interaction} \label{sec:el_ph_int}
Next, we introduce a longitudinal acoustic phonon to the system, whose low-energy dispersion as a function of momentum $\boldsymbol{q}$ is $\Omega_{\boldsymbol{q}} = v_s q$, where $v_s$ is the sound velocity and $q = |\boldsymbol{q}|$. The bare phonons are described by the Hamiltonian
\begin{equation}
	H_p=\sum_{\boldsymbol{q}}\hbar\Omega_{\boldsymbol{q}}\left(b^\dagger_{\boldsymbol{q}}b_{\boldsymbol{q}}+\frac12\right)\,,
\end{equation}
where $b_{\boldsymbol{q}}$ ($b_{\boldsymbol{q}}^\dagger$) are phononic annihilation (creation) operators. These phonons are coupled to the electrons in the system via the electron-phonon interaction \cite{bruus_manybody_2004}
\begin{equation}\label{eq:el_ph_int}
	V_\text{ep}=\sum_{\nu \nu^\prime}\sum_{\boldsymbol{q}}\gamma_{\nu\nu^\prime}(\boldsymbol{q})c_\nu^\dagger c_{\nu^\prime}\left(b_{\boldsymbol{q}} + b_{-\boldsymbol{q}}^\dagger\right)\,,
\end{equation}
such that the full Hamiltonian of the many-particle system is
\begin{equation}\label{eq:h_full}
	H=\mathcal{H}_e+H_p+V_\text{ep}\,.
\end{equation}
Given that we are interested in small-momentum physics, we neglect Umklapp scattering. The electron-phonon matrix elements then read
\begin{equation}\label{eq:coupl_elements}
	\gamma_{\nu\nu^\prime}(\boldsymbol{q})=i e\sqrt{\frac{\hbar q}{2\rho V v_s}}\braket{\nu|\hat{V}_{\boldsymbol{q}}\,e^{i\boldsymbol{q\hat{r}}}|\nu^\prime}\,,
\end{equation}
where $\hat{V}_{\boldsymbol{q}}$ is the Fourier transform of the electrostatic electron-ion potential and $\rho$ is the ion-mass density. We simplify the interaction by assuming a local deformation potential that acts trivially on the electronic orbital and spin degrees of freedom, $\hat{V}_{\boldsymbol{q}}=D\,\hat{\mathds{1}}$. The leading contribution of the electron-phonon matrix elements for small momenta is given by (see Appendix \ref{app:el_ph_me} for the arbitrary-order expression with respect to~$\boldsymbol{q}$)
\begin{equation}\label{eq:abs_sq_m_el}
	\abs{\gamma_{\nu\nu^\prime}(\boldsymbol{q})}^2 = \frac{e^2 D^2 \hbar}{2\rho V v_s}\,q\,\delta_{\nu\nu^\prime}+\mathcal{O}\left(q^2\right)\,.
\end{equation}

\subsection{Phonon self-energy}\label{sec:rpa}
Phonon propagation is described by their Matsubara Green's function
\begin{equation}
	\mathcal{D}(\boldsymbol{q};\tau)=-\braket{\hat{\mathcal{T}_\tau}A_{\boldsymbol{q}}(\tau)A_{\boldsymbol{q}}^\dagger(0)}\,,
\end{equation}
where the thermal average and the time dependence of the operators are given as described below Eq.~\eqref{eq:free_el_gf} but with respect to the full Hamiltonian $H$ from Eq.~\eqref{eq:h_full} and with $A_{\boldsymbol{q}}=b_{\boldsymbol{q}} + b_{-\boldsymbol{q}}^\dagger$. Performing a perturbative expansion of the phonon propagator with respect to the electron-phonon interaction yields the Dyson equation
\begin{equation}
	\mathcal{D}(\boldsymbol{q};i\Omega_n)^{-1}=\mathcal{D}^{(0)}(\boldsymbol{q};i\Omega_n)^{-1} - \Sigma_{\boldsymbol{q}}(i\Omega_n)\,,
\end{equation}
where $i\Omega_n = 2n\pi / (\hbar \beta)$ are bosonic Matsubara frequencies, $\Sigma_{\boldsymbol{q}}(i\Omega_n)$ is the phonon self-energy, and with the free phonon propagator \cite{bruus_manybody_2004}
\begin{equation}
	\mathcal{D}^{(0)}(\boldsymbol{q};i\Omega_n)=\frac{2\hbar\Omega_{\boldsymbol{q}}}{(i\hbar\Omega_n)^2-(\hbar\Omega_{\boldsymbol{q}})^2}\,.
\end{equation}
The electron-phonon vertex from Eq.~\eqref{eq:el_ph_int} describes that a phonon may decay into an electron-hole pair. As depicted in Fig.~\ref{fig:feynman_lattice}(a), this electron-hole pair can eventually recombine to create a new phonon. In the spirit of Zhang and Zhou \cite{zhang_quantum_2020},  we sum up the infinite ladder of such particle-hole bubbles, i.e., we perform a random-phase-like approximation. This yields
\begin{align}
		\Sigma_{\boldsymbol{q}}(i\Omega_n)=\frac{1}{\beta}\sum_{i\omega_m}\sum_{\nu\nu^\prime}\left[\vphantom{\abs{\gamma_{\nu\nu^\prime}(\boldsymbol{q})}^2}\right.
		&\abs{\gamma_{\nu\nu^\prime}(\boldsymbol{q})}^2 \mathcal{G}^{(0)}(\nu^\prime; i\omega_m)\nonumber\\
		&\times \mathcal{G}^{(0)}(\nu;i\omega_m + i\Omega_n)\left.\vphantom{\abs{\gamma_{\nu\nu^\prime}(\boldsymbol{q})}^2}\right]\,.
	\label{eq:ph_se_it}
\end{align}
Performing the sum over $\omega_n$ and the analytical continuation $i\Omega_n\rightarrow\Omega + i0^+$, and considering phonon frequencies $\Omega$ that satisfy the unperturbed dispersion yields the retarded phonon self-energy $\Sigma^r_{\boldsymbol{q}}:= \Sigma_{\boldsymbol{q}}(\Omega_{\boldsymbol{q}}+i0^+)$. Explicitly, we have
\begin{align}\label{eq:ph_se_ret}
		\Sigma^r_{\boldsymbol{q}}=\int_{-\infty}^{\infty}\mathrm{d}\omega \sum_{\nu \nu^\prime}\,&\left\{\vphantom{\frac12}\right.\hbar n_F(\hbar\omega)\abs{\gamma_{\nu\nu^\prime}(\boldsymbol{q})}^2\nonumber\\
		&\times
			\left[\vphantom{G_r^{(0)}}\right. G_r^{(0)}(\nu;\omega+\Omega_{\boldsymbol{q}})\,\rho_{\nu^\prime}(\omega)\nonumber\\
			&+\rho_\nu(\omega)\, G^{(0)}_r(\nu^\prime;\omega-\Omega_{\boldsymbol{q}})^*\left.\vphantom{G_r^{(0)}}\right]\left.\vphantom{\frac12}\right\}\,.	
\end{align}

In the remainder, we are interested in the leading-order contributions to the long-wavelength expansion. While the real part corresponds to a renormalization $\Delta v_s$ of the sound ve\-lo\-ci\-ty, $\operatorname{Re}\Sigma_{\boldsymbol{q}}=\hbar \Delta v_s q + \mathcal{O}(q^2)$, the imaginary part describes the spectral width $\Gamma_{\boldsymbol{q}} =-\operatorname{Im}\Sigma_{\boldsymbol{q}}$ and thus the attenuation $\alpha_{\boldsymbol{q}}=-\Gamma_{\boldsymbol{q}}/(\hbar v_s)$ of the phonon. Inserting the simplified electron-phonon interaction matrix elements from Eq.~\eqref{eq:abs_sq_m_el}, we deduce the final form for the retarded phonon self-energy
\begin{align}
		\Sigma^r_{\boldsymbol{q}}=\int_{-\infty}^{\infty}\mathrm{d}\omega\sum_{\nu}
			\left\{\vphantom{\frac12}\right.&\,A_\Sigma\, n_F(\hbar\omega)\rho_{\nu}(\omega)\nonumber\\
			&\times
				\left[\vphantom{G_r^{(0)}}\right.G_r^{(0)}(\nu;\omega+\Omega_{\boldsymbol{q}})\nonumber\\
				 &+ G^{(0)}_r(\nu^\prime;\omega-\Omega_{\boldsymbol{q}})^*\left.\vphantom{G_r^{(0)}}\right]\left.\vphantom{\frac12}\right\}\,,\label{eq:ph_se_ret_lwl}
\end{align}
where $A_\Sigma= e^2D^2\hbar^2 q/(2\rho V v_s)$.

\section{Results and experimental comparison}\label{sec:results}
\subsection{Phonon-velocity renormalization} \label{sec:dv}
A first important experimental observable is the phonon-velocity renormalization $\Delta v_s$ as a function of magnetic-field strength. Taking the real part of the phonon self-energy from Eq.~\eqref{eq:ph_se_ret_lwl} and only considering the leading-order term in $q$ yields
\begin{align}
			\Delta v_s = \frac{e^2 D^2 }{\rho V v_s}\int_{-\infty}^{\infty}\mathrm{d}\omega \sum_\nu\left[\vphantom{G_\text{r}^{(0)}}\right. &\hbar n_\text{F}(\hbar\omega) \rho_\nu(\omega)\nonumber\\ &\times \operatorname{Re}G_\text{r}^{(0)}(\nu;\omega)\left.\vphantom{G_\text{r}^{(0)}}\right]\,.\label{eq:dv_finite_temp}
\end{align}
A more insightful way to write the phonon-velocity renormalization is obtained by considering the zero-temperature limit, $T=0$. In this limit, we can perform the $\omega$ integration analytically. We then find
\begin{align}
	\Delta v_s 	&= - A_v B v_s^2 \hbar\int_{-\infty}^\infty\frac{\mathrm{d}k_z}{2\pi}\sum_s 
		\left[\vphantom{\sum_t}\right.	\rho_{0sk_z}(0)+ \sum_{nt}\rho_{nstk_z}(0)\left.\vphantom{\sum_t}\right]\nonumber \\
				&=- A_v B v_s^2 \hbar\, \mathcal{D}(\mu)\,,\label{eq:dv_zero_temp}
\end{align}
where $\mathcal{D}(\mu)=\int_{-\infty}^\infty\frac{\mathrm{d}k_z}{2\pi}\sum_s \left[\rho_{0sk_z}(0) + \sum_{nt}\rho_{nstk_z}(0)\right]$ is the density of states at the Fermi level and $A_v=e^3D^2/(4\pi\rho v_s^3\hbar^2)$. Eq.~\eqref{eq:dv_zero_temp} of a natural generalization to Eq.~(16) in the work of Zhang and Zhou \cite{zhang_quantum_2020} to a system with finite Landau-level broadening. In the limit of delta-shaped spectral functions, $\Gamma\rightarrow0^+$, and if the Fermi level approaches the band bottom of the $\nu$-th Landau band, a van Hove singularity occurs, since $\Delta v_s \propto 1 / \sqrt{\mu - \epsilon_\nu(k_z = 0)}$. The impurity-induced Landau-band broadening $\Gamma > 0$ regularizes these singularities and yields smooth quantum oscillations of the phonon-velocity renormalization. 

\begin{figure}
	\centering 
	\includegraphics[width=\linewidth]{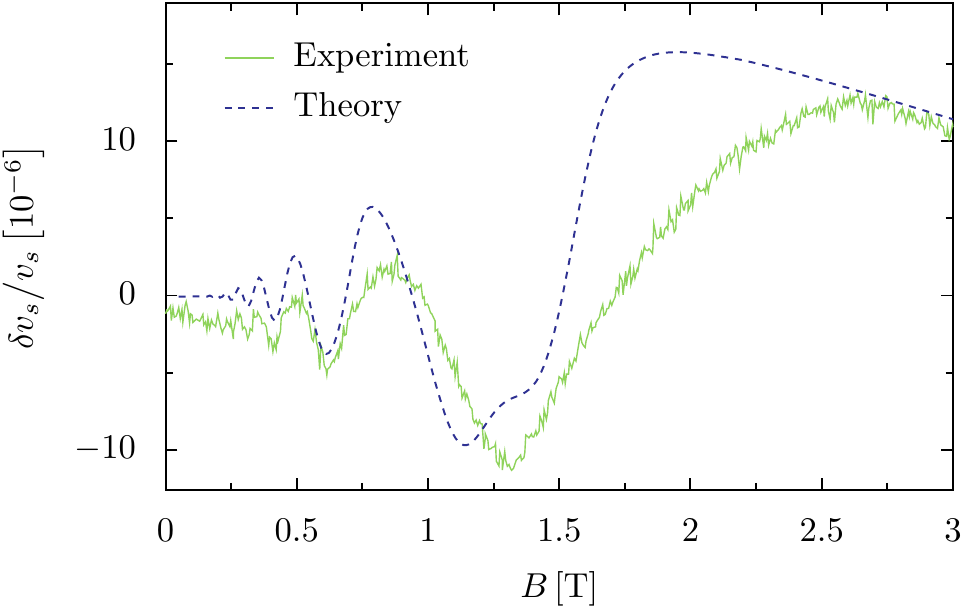} 
	\caption{Relative phonon-velocity renormalization as a function of the applied magnetic field with fixed charge-carrier density. The blue, dashed line is obtained from Eq.~\eqref{eq:dv_finite_temp}. The green, solid line is experimental data taken from Ref. \cite{galeski_origin_2021}. The parameters are given by: $m=\SI{1}{\milli\eV},v_z=\SI{15340}{\m\per\s},\,g=18,\sqrt{2e\hbar |v_x|v_y}=\SI{12.5}{\milli\eV\tesla\tothe{-\frac12}}, v_s=\SI{3375}{\meter\per\second}, n_0=\SI{2e-7}{\per\cubic\angstrom},T=\SI{2}{\kelvin},\Gamma=\SI{0.5}{\milli\electronvolt},\Lambda=\SI{25}{\milli\electronvolt}, A_v = \SI{78.54}{\per\mega\tesla}\,.$} 
	\label{fig:dv} 
\end{figure}

To assess the accuracy of our theoretical prediction in Eq.~\eqref{eq:dv_finite_temp}, we compare it with experimental data taken from Ref.~\cite{galeski_origin_2021}, which investigated longitudinal acoustic modes along the $x$ axis, $\boldsymbol{q} = q \,\boldsymbol{e}_x$. The field dependence of the experimentally observed sound velocity $\tilde{v}_s(B)$ is handed down to the phonons from the electrons via the electron-phonon coupling. Roughly speaking, the latter renormalizes the sound velocity by both an oscillatory term and a constant offset. Since the experiment measures the change of the sound velocity with respect to the value at a small reference field, the constant offset cannot be extracted from the experimental data. We thus define
\begin{equation}\label{eq:exp_delta_vs}
	\delta v_s(B) = \tilde{v}_s(B) - \tilde{v}_s(B_\text{ref}) = \Delta v_s(B) - \Delta v_s(B_{\text{ref}})\,,
\end{equation}
where $B_\text{ref}$ is the reference magnetic field used in the experiments ($|B_\text{ref}| = \SI{0.0339}{\tesla} \ll B_\text{QL}$). For the theoretical prediction, we estimate the value of the sound-velocity renormalization at the reference field $B_\text{ref}$ by averaging over the quantum oscillations in $\Delta v_s(B)$ for small magnetic fields $\SI{0.05}{\tesla}<B<\SI{0.3}{\tesla}\,$. The microscopic parameters for the electronic part of our Hamiltonian \eqref{eq:h_finite_field}, which determine the peak structure of the quantum oscillations, are chosen in a similar range as the values used in Refs.~\cite{chen_magnetoinfrared_spec_2015, galeski_origin_2021} for the description of transport experiments. Since the ultrasound measurements have been performed on different samples, however, we allow for slight modifications of the parameter values. In the remainder, we use the following set of electronic parameters: $m=\SI{1}{\milli\eV},\,v_z=\SI{15340}{\m\per\s},\,g=18,\,\sqrt{2e\hbar|v_x|v_y}=\SI{12.5}{\milli\eV\tesla\tothe{-\frac12}},\,n_0=\SI{2e-7}{\per\cubic\angstrom}$.  In addition, our model features the parameters of the phononic subsystem, and of the electron-phonon coupling. The sound velocity has been taken from experimental data from Ref.~\cite{galeski_origin_2021}, $v_s=\SI{3375}{\meter\per\second}$. The global prefactors of the phonon-velocity renormalization and attenuation in Eqs.~\eqref{eq:dv_finite_temp} and \eqref{eq:att_powerlaw}, $A_v$ and $A_\alpha=e^3D^2q^2/(4\rho v_s^2\hbar^3)$ (see Eq. \ref{eq:alpha_zero_temp} below), respectively, which determine the amplitude of the quantum oscillations, allow us to fit the phononic parameters for a magnetic-field range from zero up to fields slightly beyond the quantum-limit regime, $B\le\SI{3}{\tesla}$, see Fig. \ref{fig:dv}.

For the considered sample, we obtain the fit parameters $A_v = \SI{78.54}{\per\mega\tesla}$ and $A_\alpha = \SI{0.942}{\milli\electronvolt\decibel\per\tesla\per\centi\meter}$, which suggest that the deformation potential is weak compared to other materials. This is consistent with experimental observations from time-resolved optical-reflectivity measurements \cite{zhang_ultrafast_2019}. Furthermore, materials with stronger electron-phonon interaction, such as \ce{TbTe3} \cite{saintpaul_elastic_2016}, typically show quantum oscillations in phonon dynamics that are several orders of magnitude larger than the ones for \ce{ZrTe5}. The smallness of the electron-phonon coupling in the gapless Dirac semimetal scenario is also consistent with our initial assumption that the material does not exhibit charge-density-wave order of the electrons, in particular also not induced by phonon-mediated electron-electron interactions \cite{Qin_cdw_2020}.

\subsection{Sound attenuation} \label{sec:att}
In addition to the phonon-velocity renormalization, the decay of the ultrasound signal that is observed in the experiments directly allows to extract the sound-attenuation factor $\alpha_{\boldsymbol{q}}$. To compare the experimental results with our model, we expand Eq.~\eqref{eq:ph_se_ret_lwl} to second order in $q$. This yields
\begin{align}\label{eq:alpha_finite_temp}
	\alpha_{\boldsymbol{q}}&=\pi\frac{e^2 D^2}{2\rho V v_s \hbar}\,q^2\int_{-\infty}^{\infty}\mathrm{d}\omega\,\hbar n^\prime_\text{F}(\hbar \omega)\sum_\nu \rho_\nu(\omega)^2.
\end{align}
At zero temperature, where $n^\prime_\text{F}(\epsilon)=-\delta(\epsilon)$, the sound attenuation becomes
\begin{align}
		\alpha_{\boldsymbol{q}}=- A_\alpha B v_s \hbar\int_{-\infty}^\infty \frac{\mathrm{d}k_z}{2\pi}\sum_s
			&\left[\vphantom{\sum_t}\right.	\rho_{0sk_z}(0)^2 \nonumber\\
			&+\sum_{nt}\rho_{nstk_z}(0)^2\left.\vphantom{\sum_t}\right]\,,	\label{eq:alpha_zero_temp}
\end{align}
where $A_\alpha=e^3D^2q^2/(4\rho v_s^2\hbar^3)$ as mentioned above.

\begin{figure}
	\centering 
	\includegraphics[width=\linewidth]{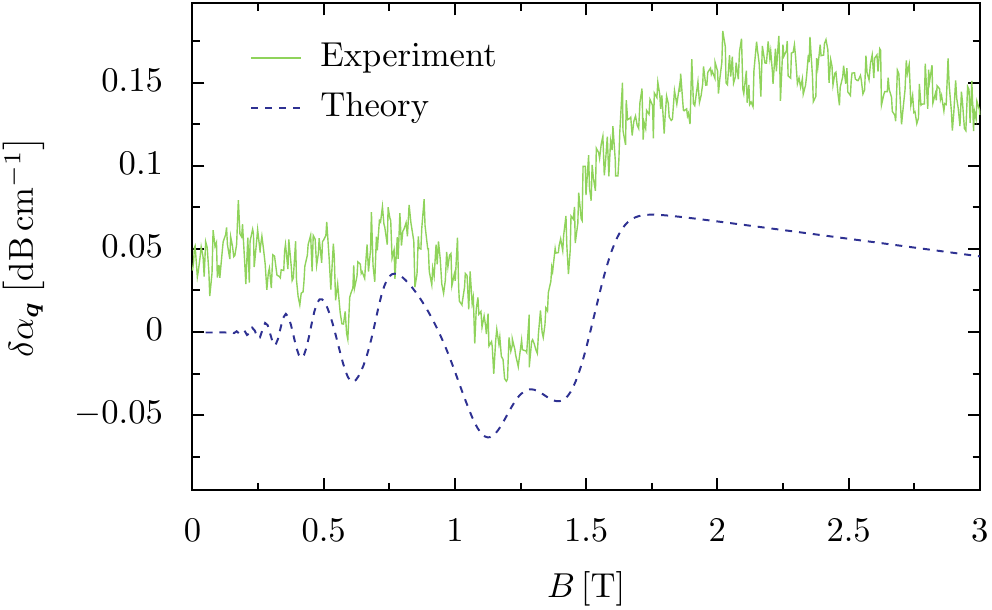} 
	\caption{Sound attenuation as a function of the applied magnetic field with fixed charge-carrier density. The blue, dashed line is obtained from Eq.~\eqref{eq:alpha_finite_temp}. The green, solid line is experimental data taken from Ref. \cite{galeski_origin_2021}. The parameters are given by: $m=\SI{1}{\milli\eV},v_z=\SI{15340}{\m\per\s},g=18,\sqrt{2e\hbar |v_x|v_y}=\SI{12.5}{\milli\eV\tesla\tothe{-\frac12}},v_s=\SI{3375}{\meter\per\second}, n_0=\SI{2e-7}{\per\cubic\angstrom},T=\SI{2}{\kelvin},\Gamma=\SI{0.5}{\milli\electronvolt},\Lambda=\SI{25}{\milli\electronvolt}, A_\alpha = \SI{0.942}{\milli\electronvolt\decibel\per\tesla\per\centi\meter}\,.$} 
	\label{fig:alpha} 
\end{figure}

As for the phonon-velocity renormalization, the electron-phonon coupling induces quantum oscillations in the sound attenuation, which are, similarly to Eq. \eqref{eq:exp_delta_vs}, measured with respect to the reference-field value
\begin{equation}
	\delta \alpha_{\boldsymbol{q}}(B) = \alpha_{\boldsymbol{q}}(B) - \alpha_{\boldsymbol{q}}(B_\text{ref})\,.
\end{equation} As for the sound-velocity renormalization, we obtain $\alpha_{\boldsymbol{q}}(B_\text{ref})$ by averaging over $\alpha_{\boldsymbol{q}}(B)$ for small fields $\SI{0.05}{\tesla}<B<\SI{0.3}{\tesla}\,$. For the same set of parameters that has been used for the data shown in Fig.~\ref{fig:dv}, we obtain a quantitatively good agreement of Eq.~\eqref{eq:alpha_finite_temp} with experimental data from Ref.~\cite{galeski_origin_2021} up to the quantum-limit regime, as shown in Fig. \ref{fig:alpha}.

Combining our theoretical results for the  phonon velocity renormalization and the sound attenuation, we find that the gapless Dirac model from Eq.~\eqref{eq:h_finite_field} supplemented by a simple electron-phonon coupling building on the deformation potential in Eq.~\eqref{eq:abs_sq_m_el} is able to describe experimental observables from phonon dynamics in \ce{ZrTe5} on a quantitative level up to the quantum-limit regime.  Note that the very same model is also consistent with measurements of electric and thermoelectric transport quantities \cite{galeski_origin_2021}.

\section{Effect of a charge-density-wave gap} \label{sec:cdw}

\subsection{Density-density interaction and charge-density-wave state}
As discussed in Sec.~\ref{sec:intro}, an alternative explanation for the plateau-like features in the Hall resistivity of \ce{ZrTe5} is a bulk CDW whose ordering wave vector spans the Dirac pocket along the direction of the applied magnetic field. In the following, we discuss what effect a CDW gap would have on ultrasound measurements.

In general, different mechanisms might lead to the formation of a CDW in \ce{ZrTe5}, including direct electron-electron interactions, or electronic interactions deriving from electron-phonon coupling \cite{Qin_cdw_2020}. For the sake of our discussion, the microscopic mechanism leading to the formation of a CDW is not important, and we consider a possibly effective (e.g., phonon-mediated) electronic density-density interaction in the local approximation
\begin{align}
	V_\text{ee}	&=\frac{g_0}{2}\int\mathrm{d}^3 r\, n(\boldsymbol{r})^2 \\
				&= 
				\begin{aligned}[t]
					\frac{g_0}{2}\int\mathrm{d}^3 r\sum_{\tau\tau^\prime}\sum_{\sigma \sigma^\prime}\, 	&\psi_{\tau\sigma}(\boldsymbol{r})^\dagger\, \psi_{\tau\sigma}(\boldsymbol{r})\\
																										&\times \psi_{\tau^\prime\sigma^\prime}(\boldsymbol{r})^\dagger\, \psi_{\tau^\prime\sigma^\prime}(\boldsymbol{r})\,,
				\end{aligned}
\end{align}
where $n(\boldsymbol{r}) = \sum_{\tau\sigma}\psi_{\tau\sigma}(\boldsymbol{r})^\dagger \psi_{\tau\sigma}(\boldsymbol{r})$ is the electron density and $g_0$ is the coupling strength. In a mean-field-like spirit, we replace this interaction by the effective CDW Hamiltonian
\begin{align}\label{eq:h_mf_rs}
	H_M&=H_e + V_M\\
				&=\int\mathrm{d}^3 r\,\psi(\boldsymbol{r})^\dagger\left[h_\text{e}(\boldsymbol{r})+g_0\,\bar{n}(\boldsymbol{r})\,\mathds{1}\right]\psi(\boldsymbol{r})\,.
\end{align}
We assume $\bar{n}$ to exhibit periodic modulations along the $z$ axis with a single wave number $Q$, i.e., $\bar{n}(\boldsymbol{r})=2\,\bar{n}_Q\cos(Qz)$, which implies the real-space periodicity $\lambda = 2\pi/Q$. The latter is in general much larger than the lattice constant $a_z$ along the $z$ axis, $\lambda \gg a_z$. 

However, instead of calculating the CDW amplitude $\bar{n}_Q$ self-consistently (as in Ref. \cite{Qin_cdw_2020}), it is in our case sufficient to consider it a parameter, since we are merely interested in a qualitative analysis of how a CDW gap affects the phonon self-energies, and thus ultrasound measurements. This is sufficient to compare with experiments, since $\bar{n}_Q$ is directly related to the CDW-induced gap in the electronic structure, and can thus be measured experimentally. 

Expressing Eq.~\eqref{eq:h_mf_rs} in terms of the Landau-band eigenmodes, given in Eqs.~\eqref{eq:zero_modes} and \eqref{eq:higher_order_modes}, yields $V_M = V_0 + \sum_{n=1}^{\infty}V_n\,,$ with the contribution from the zeroth Landau bands
\begin{align}
		V_0 =\sum_{k_xk_z}\sum_{ss^\prime}\left[\vphantom{c_{k_x(k_z+Q)0s}^\dagger}\right.&\Delta_{ss^\prime}(k_z + Q, k_z) c_{k_x(k_z+Q)0s}^\dagger c_{k_xk_z0s^\prime}\nonumber \\
												&+ \mathrm{h.c.}\left.\vphantom{c_{k_x(k_z+Q)0s}^\dagger}\right]\,,\label{eq:v_mf}
\end{align}
where 
\begin{equation}
	\Delta_{ss^\prime}(k_z + Q, k_z) = g_0\,\bar{n}_Q\braket{\boldsymbol{w}_{0s}(k_z+Q)|\boldsymbol{w}_{0s^\prime}(k_z)}
\end{equation}
is the overlap of the corresponding eigenvectors from Eq.~\eqref{eq:eigenvectors_0}, and the potential arising from the $n$-th Landau bands reads as
\begin{align}
	V_n=\sum_{k_xk_z}\sum_{\substack{ss^\prime \\ tt^\prime}}
	\left[\vphantom{c_{k_x(k_z+Q)0s}^\dagger}\right.&\Delta_{nsts^\prime t^\prime}(k_z + Q, k_z)\nonumber \\
													& \times c_{k_x(k_z+Q)nst}^\dagger c_{k_xk_zns^\prime t^\prime} + \mathrm{h.c.}\left.\vphantom{c_{k_x(k_z+Q)0s}^\dagger}\right]\,,
\end{align} 
where
\begin{align}
		\Delta_{nsts^\prime t^\prime}(k_z + Q, k_z) &= 	g_0\,\bar{n}_Q\nonumber\\
														&\times\braket{\boldsymbol{w}_{nst}(k_z+Q)|\boldsymbol{w}_{ns^\prime t^\prime}(k_z)}
\end{align}
is the overlap of the eigenvectors from Eq.~\eqref{eq:eigenvectors_n}. This term mixes electronic states with different wave numbers $k_z$ and Landau-band indices $s$ and $t$. In the quantum-limit regime, the low-energy physics is dominated by the zeroth Landau bands. We, therefore, project out all higher Landau bands. Motivated by the scenario proposed in Refs.~\cite{Qin_cdw_2020, tang_three_dimensional_2019}, we consider $Q = 2 k_{F,z}$, and fold the Hamiltonian back into the reduced Brillouin zone  using $k_z = \kappa_z + \xi Q$  with $|\kappa_z|<k_{F,z}$ and integer $\xi$. The Hamiltonian thus becomes
\begin{align}
	\begin{split}
		H_m=\sum_{k_x\kappa_z}c_{k_x\kappa_z}^\dagger h_m(\kappa_z)\,c_{k_x\kappa_z}\,,
	\end{split}
\end{align}
where we have introduced the fermionic spinors $c_{k_x\kappa_z}=(c_{k_x(\kappa_z+\xi Q)0s})^T$ that contain the modes from the particle and hole branches of the zeroth Landau bands, $s=\pm$, and that have wave numbers $k_z$ differing from $\kappa_z$ by integer multiples $\xi$ of the CDW wave number $Q$. Furthermore, we have defined the effective Hamiltonian 
\begin{align}
		(h_m(\kappa_z))_{\xi s \xi^\prime s^\prime}&=	\epsilon_{0s}(\kappa_z+\xi Q)\,\delta_{\xi\xi^\prime}\delta_{s s^\prime} \nonumber\\
																&+ \Delta_{ss^\prime}(\kappa_z + \xi Q, \kappa_z + \xi^\prime Q)\delta_{\xi(\xi^\prime+1)\nonumber}\\
																&+ \Delta_{s^\prime s}(\kappa_z + \xi^\prime Q, \kappa_z + \xi Q)\, \delta_{(\xi+1)\xi^\prime}\,.\label{eq:h_mf}
\end{align}
Because the system is inversion symmetric, it suffices to consider $\kappa_z > 0$. The relevant bands are then $\xi = -1$ and $\xi=0$, and the single-particle Hamiltonian from Eq.~\eqref{eq:h_mf} becomes
\begin{align}
		h_m(\kappa_z)=\,	&E_+(\kappa_z)\,\tau_0\sigma_3 + E_-(\kappa_z)\,\tau_3\tau_3 \nonumber\\
								&+ \Delta_g(\kappa_z)\,\tau_1\sigma_3+\Delta_\text{ph}(\kappa_z)\,\tau_1\sigma_1\,,\label{eq:h_mf_reduced}
\end{align}
where $E_\pm(\kappa_z) = [\epsilon_{0+}(\kappa_z-Q)\pm\epsilon_{0+}(\kappa_z)]/2\,$, and with
\begin{align}
		\Delta_g(\kappa_z) = \frac{g_0\bar{n}_Q}{2}&\left\{\vphantom{\sqrt{\frac12}}\right.
			\sqrt{1 + \frac{m_*}{\epsilon_{0+}(\kappa_z - Q)}}\nonumber\\
			&\times\sqrt{1 + \frac{m_*}{\epsilon_{0+}(\kappa_z)}}\nonumber
\\
		&-
			\sqrt{1 - \frac{m_*}{\epsilon_{0+}(\kappa_z - Q)}}\nonumber\\
			&\times\sqrt{1 - \frac{m_*}{\epsilon_{0+}(\kappa_z)}}\left.\vphantom{\sqrt{\frac12}}\right\}\,,
\end{align}
and
\begin{align}
		\Delta_\text{ph}(\kappa_z) = -\frac{g_0\bar{n}_Q}{2}&\left\{\vphantom{\sqrt{\frac12}}\right.
			\sqrt{1 + \frac{m_*}{\epsilon_{0+}(\kappa_z - Q)}}\nonumber\\
			&\times \sqrt{1 - \frac{m_*}{\epsilon_{0+}(\kappa_z)}}\nonumber\\
		&+
			\sqrt{1 - \frac{m_*}{\epsilon_{0+}(\kappa_z - Q)}}\nonumber\\ 
			&\times\sqrt{1 + \frac{m_*}{\epsilon_{0+}(\kappa_z)}}\left.\vphantom{\sqrt{\frac12}}\right\}\,.
\end{align}
Here, the matrix element $\Delta_g(k_{F,z})$ describes the coupling of modes within the same unperturbed Landau band at the boundary of the reduced Brillouin zone, and thus the size of the CDW-induced gap, whereas $\Delta_\text{ph}(k_{F,z})$ couples the particle and the hole branches of the zeroth Landau levels. Diagonalizing Eq.~\eqref{eq:h_mf_reduced} yields the final expression for the band structure in the CDW scenario
\begin{align}
		E_{st}(\kappa_z)=s\left\{\vphantom{\sqrt{E^2}}\right.	&E_+(\kappa_z)^2 + E_-(\kappa_z)^2\nonumber\\
																			&+\Delta_\text{ph}(\kappa_z)^2+\Delta_g(\kappa_z)^2\nonumber\\
		&+2t
			\left[\vphantom{\Delta_g(\kappa_z)^2}\right.E_-(\kappa_z)^2\Delta_\text{ph}(\kappa_z)^2\nonumber\\
													&+E_+(\kappa_z)^2 E_-(\kappa_z)^2\nonumber \\
													&+E_+(\kappa_z)^2 \Delta_g(\kappa_z)^2 \left.\vphantom{\Delta_g(\kappa_z)^2}\right]^\frac12\left.\vphantom{\sqrt{E^2}}\right\}^\frac12\,,\label{eq:lbs_cdw}
\end{align}
where $s,t = \pm$. We observe that in our simple model the electrons remain gapless even in the presence of a CDW unless there is a finite effective Dirac gap $m_* \neq 0$. This can be traced back to the orthogonality of the electronic states in that limit, see Eq.~\eqref{eq:eigenvectors_0}, which results in $\Delta_g=0$. Physically, this corresponds to the fact that a spinless CDW cannot couple electronic states of orthogonal spins. For any finite Dirac gap, however, the CDW induces a gap $M > 0$ between the electronic bands at the Fermi wave vector. The density of states at the Fermi level is reduced drastically if the gap is larger than the level broadening. As an aside, we note that the gap $M > 0$ ceases to grow at very large fields due to a crossing between the Landau bands of the conduction and valence bands.

\begin{figure}
	\centering
	\includegraphics[width=\linewidth]{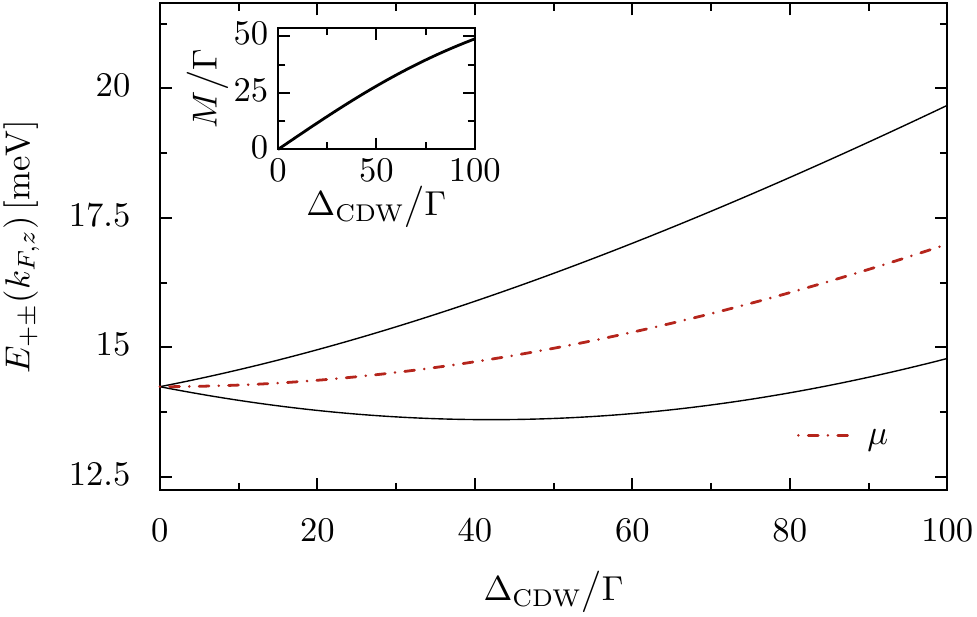}
	\caption{The positive branches of the band structure in the charge-density-wave scenario (black, solid) from Eq.~\eqref{eq:lbs_cdw} at the boundary of the first Brillouin zone $\kappa_z=k_{F,z}$ and the Fermi level (red, dash-dotted) as a function of $\Delta_\text{CDW}=g_0\bar{n}_Q$. The inset shows the CDW-induced gap between the positive electronic bands at the Fermi wave vector as a function of $\Delta_\text{CDW}$. The parameters are given by: $m=\SI{1}{\milli\eV},\,v_z=\SI{15340}{\m\per\s},\,g=18,\,\sqrt{2e\hbar |v_x|v_y}=\SI{12.5}{\milli\eV\tesla\tothe{-\frac12}},\,\mu(\Delta_{CDW}=0)=\SI{14.239}{\milli\electronvolt},\, k_{F,z}=\SI{0.135}{\per\angstrom}\,,T=\SI{0}{\kelvin},\,\Gamma=\SI{0.1}{\milli\electronvolt},\, B=\SI{10}{\tesla},\, \Lambda=\SI{25}{\milli\electronvolt}.$}
	\label{fig:lbs_cdw}
\end{figure}

\subsection{Charge-density-wave induced decay of phonon self-energy} \label{sec:cdw_decay}
\begin{figure}
	\centering
	\includegraphics[width=\columnwidth]{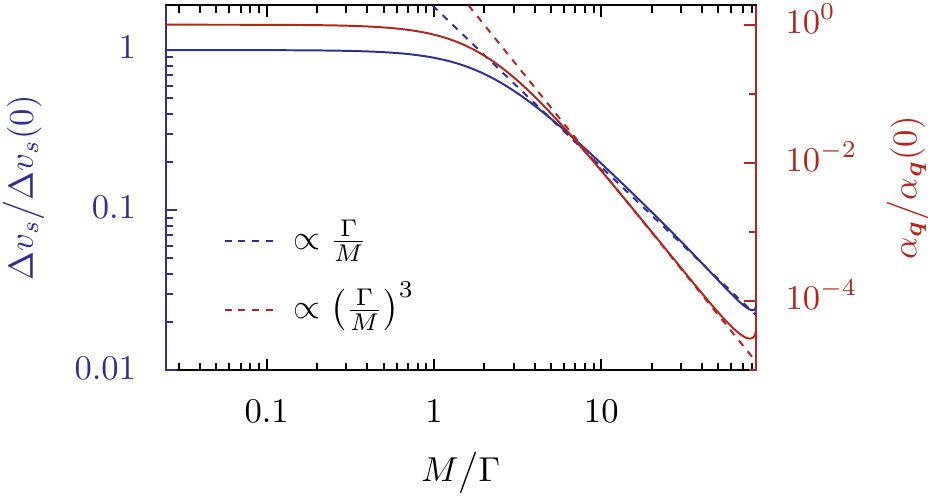}
	\caption{Phonon-velocity renormalization (blue, solid) and sound attenuation (red, solid) in the charge-density-wave scenario relative to the Dirac-metal scenario as a function of the gap. If the gap is larger than the level broadening $\Gamma$, both quantities decay with a power law [see Eqs.~\eqref{eq:dv_powerlaw} (blue, dashed) and \eqref{eq:att_powerlaw} (red, dashed), respectively]. The parameters are given in Fig. \ref{fig:lbs_cdw}.}
	\label{fig:dv_att_cdw}
\end{figure}

We close our discussion by analyzing the effect of a CDW on the phonon-velocity re\-nor\-ma\-li\-za\-tion and the sound attenuation.  Due to the reduced density of states at the Fermi level resulting from the opening of a CDW gap, we observe a power-law decay of the phonon-velocity renormalization and the sound attenuation compared to the gapless Dirac-metal state whenever the gap at the Fermi level is large compared to the level broadening, see Fig.~\ref{fig:dv_att_cdw}. The phonon-velocity renormalization and the sound attenuation scale as
\begin{equation}\label{eq:dv_powerlaw}
	\frac{\Delta v_s(\Delta_\text{CDW})}{\Delta v_s(0)} \propto \frac{\Gamma}{M}
\end{equation} 
and 
\begin{equation}\label{eq:att_powerlaw}
	\frac{\alpha_{\boldsymbol{q}}(\Delta_\text{CDW})}{\alpha_{\boldsymbol{q}}(0)} \propto \left(\frac{\Gamma}{M}\right)^3\,,
\end{equation}
respectively.

Our model thus predicts that a CDW, via the related gap, induces a significant suppression in the amplitude of quantum oscillations in phonon-propagation-related properties. This result can for example be compared with the proposition of Ref.~\cite{Qin_cdw_2020} that the CDW-induced gap in \ce{ZrTe5} in the quantum-limit regime may be of the order of several $\SI{10}{\milli\electronvolt}$, and thus  large compared to the Landau-level broadening assumed in the present work. Our calculation then suggests that the sound attenuation might be suppressed by a factor of $10^{-4}$ as compared to the gapless Dirac scenario. To still explain the experimentally observed field dependence of phonon properties, the deformation potential $D$ would have to be rescaled by a factor $10^2$, see Eq.~\eqref{eq:alpha_finite_temp}. A CDW in \ce{ZrTe5} would thus require a significant electron-phonon interaction, which is not expected in this compound \cite{zhang_ultrafast_2019}. Thus, while the CDW scenario in principle is possible, we believe that the metallic Dirac model, which consistently explains a number of independent measurements in addition to the presently discussed ultrasound properties \cite{galeski_origin_2021} seems much more plausible.

\section{Summary and outlook}
Starting from a microscopic description of the material, we constructed a simple low-energy model for the phonon dynamics of \ce{ZrTe5}. We showed that our model is able to describe  experimental data of ultrasound measurements (sound-velocity renormalization and sound attenuation) in \ce{ZrTe5} on a quantitative level up to the quantum-limit regime. Our results are consistent with the hypothesis that up to the quantum-limit regime, the low-energy physics of \ce{ZrTe5} is dominated by a gapless Dirac-type Fermi surface close to the $\Gamma$ point. We contrasted these results with the possible formation of a CDW. The latter would suppress the electronic density of states at the Fermi level, and thus the magnetic-field dependence of the phonon self-energy. A CDW would, therefore, only be consistent with the experimentally observed ultrasound properties if the electron-phonon coupling would be unusually large. This leads us to conclude that ultrasound measurements strongly indicate a gapless state in \ce{ZrTe5}.

The model that we developed is easily generalizable to materials, whose electronic low-energy physics are dominated by small Fermi surface pockets, such as \ce{HfTe5} \cite{galeski_unconventional_2020} or \ce{InAs} \cite{wawrzynczak_three_2021}. Future work can furthermore extend our analysis in a number of directions. One important question concerns the  strong simplification we made for the impurity-induced level broadening, which should be compared to more realistic treatments, e.g., using a self-consistent Born approximation \cite{koenye_magnetoresistance_2018, xiao_magnetoconductivity_2017, klier_transversal_2017}. Another important direction is the possible existence of additional Fermi-surface pockets that might especially be important at magnetic fields larger than the quantum limit \cite{zhang_electronic_2017, wang_thermodynamically_2021}. 

\section*{Acknowledgments}
We acknowledge support from the Deutsche Forschungsgemeinschaft (DFG)
through the Emmy Noether Programme ME4844/1-1, the Collaborative
Research Center SFB 1143, the W\"{u}rzburg-Dresden Cluster of Excellence on Complexity and
Topology in Quantum Matter--$ct.qmat$ (EXC 2147),
as well as the support of the HLD at HZDR, member of the European Magnetic
Field Laboratory (EMFL), and thank the Center for Information Services and HPC (ZIH) at TU Dresden for providing computing time.

\appendix
\section{Eigenspinors of the Landau-band Hamiltonian} \label{app:eigenstates}
Here, we discuss the eigenspinors of the Landau band Hamiltonian \eqref{eq:h_finite_field}. For the zeroth Landau bands, these read
\begin{equation} \label{eq:zero_modes}
	\begin{split}
		\braket{\boldsymbol{r}|k_xk_z0s}=\frac{1}{\sqrt{L_x L_z}}e^{i\left(xk_x+zk_z\right)} &\varphi_0(y-y_0(k_x)) \\
		&\times \begin{pmatrix} w^{+\uparrow}_{0s}(k_z) \\ 0 \\ 0 \\ w^{-\downarrow}_{0s}(k_z)) \end{pmatrix}\,,
	\end{split}
\end{equation}
while we obtain
\begin{equation} \label{eq:higher_order_modes}
	\begin{split}
		\braket{\boldsymbol{r}|k_xk_znst}=	&\frac{1}{\sqrt{L_x L_z}}e^{i\left(xk_x+zk_z\right)} \\
											&\times\begin{pmatrix} w^{+\uparrow}_{nst}(k_z)\,\varphi_n(y-y_0(k_x)) \\ w^{+\downarrow}_{nst}(k_z)\,\varphi_{n-1}(y-y_0(k_x)) \\ w^{-\uparrow}_{nst}(k_z)\,\varphi_{n-1}(y-y_0(k_x)) \\ w^{-\downarrow}_{nst}(k_z)\,\varphi_n(y-y_0(k_x)) \end{pmatrix}
	\end{split}
\end{equation}
for the higher Landau bands. Here, $L_z$ is the length of the system in $z$ direction, $y_0(k_x) = \hbar k_x / (eB)$ is the guiding center position and $\varphi_n$ is the $n$-th harmonic-oscillator eigenstate,
\begin{equation}\label{eq:eigenstate_ho}
	\begin{split}
		\varphi_n(y)= \frac{1}{\sqrt{2^nn!\sqrt{\pi}\mathcal{L}_B}}	&\exp\left[-\frac12 \left(\frac{y}{\mathcal{L}_B}\right)^2\right]\\
		&\times H_n\left(\frac{y}{\mathcal{L}_B}\right)\,,
	\end{split}
\end{equation}
with $\mathcal{L}_B = \ell_B \sqrt{v_y/|v_x|}$, and $H_n$ denoting the $n$-th Hermite polynomial. The coefficients $w^{+\uparrow}_{nst}(k_z), \dots, w^{-\downarrow}_{nst}(k_z)$ in Eqs.~\eqref{eq:zero_modes} and \eqref{eq:higher_order_modes} are the components of the eigenvectors of the block-diagonal Hamiltonian. 

For the zeroth Landau bands, the explicit form of the eigenvectors is
\begin{equation}\label{eq:eigenvectors_0}
	\begin{split}
		\boldsymbol{w}_{0s}(k_z)&:= 
		\begin{pmatrix}
			w^{+\uparrow}_{0s}(k_z) \\ 0 \\ 0 \\ w^{-\downarrow}_{0s}(k_z)
		\end{pmatrix}\\
		&=\frac{1}{\sqrt2}
		\begin{pmatrix}
			s\sqrt{1 + sm^* / \epsilon_{0+}(k_z)} \\ 0 \\ 0 \\ \operatorname{sgn}(v_zk_z)\sqrt{1 - sm^* / \epsilon_{0+}(k_z)}
		\end{pmatrix}\,.
	\end{split}
\end{equation}
For $n=1,2,\dots$, the eigenvectors are a bit more involved. The corresponding Bloch Hamiltonian in the Landau-band basis takes the form
\begin{align}\label{eq:matrix_hamiltonian}
	h_e(n,k_z):=\,&m\,\tau_3\sigma_0 + \sqrt{2e\hbar B |v_x|v_y n}\,\tau_1\sigma_3 \\
	&- \frac{1}{2} g \mu_B B\, \tau_0\sigma_3 + \hbar v_z k_z\, \tau_1 \sigma_1\,.
\end{align}

Unnormalized eigenvectors of $h_e(n, k_z)$ to the eigenvalue $\epsilon_{nst}(k_z)$ can be obtained by projecting an arbitrary vector $\boldsymbol{x}\neq 0$ onto the corresponding eigenspace via:
\begin{equation}
	\begin{split}
		\boldsymbol{\tilde{w}}_{nst}(k_z):=\, 	&\left[h_e(n, k_z) + \epsilon_{nst}(k_z)\right]\\
												&\times\left[h_e(n, k_z)^2 + \epsilon_{nst}(k_z)^2 - 2\gamma\right]\boldsymbol{x}\,,
	\end{split}
\end{equation}
where $\gamma = m^2 + 2e\hbar B |v_x|v_y n +(\hbar v_z k_z)^2 + (g \mu_B B / 2)^2$. Normalization yields the coefficients from Eq.~\eqref{eq:higher_order_modes}
\begin{equation}\label{eq:eigenvectors_n}
	\boldsymbol{w}_{nst}(k_z):=
	\begin{pmatrix}
		w^{+\uparrow}_{nst}(k_z) \\ w^{+\downarrow}_{nst}(k_z) \\ w^{-\uparrow}_{nst}(k_z) \\ w^{-\downarrow}_{nst}(k_z)
	\end{pmatrix}
	=\frac{\boldsymbol{\tilde{w}}_{nst}(k_z)}{\abs{\boldsymbol{\tilde{w}}_{nst}(k_z)}}\,.
\end{equation}
\hspace{1 mm}
\section{High-energy cut-off independence of observables} \label{app:cut-off}
\begin{figure}
	\centering
	\includegraphics[width=\columnwidth]{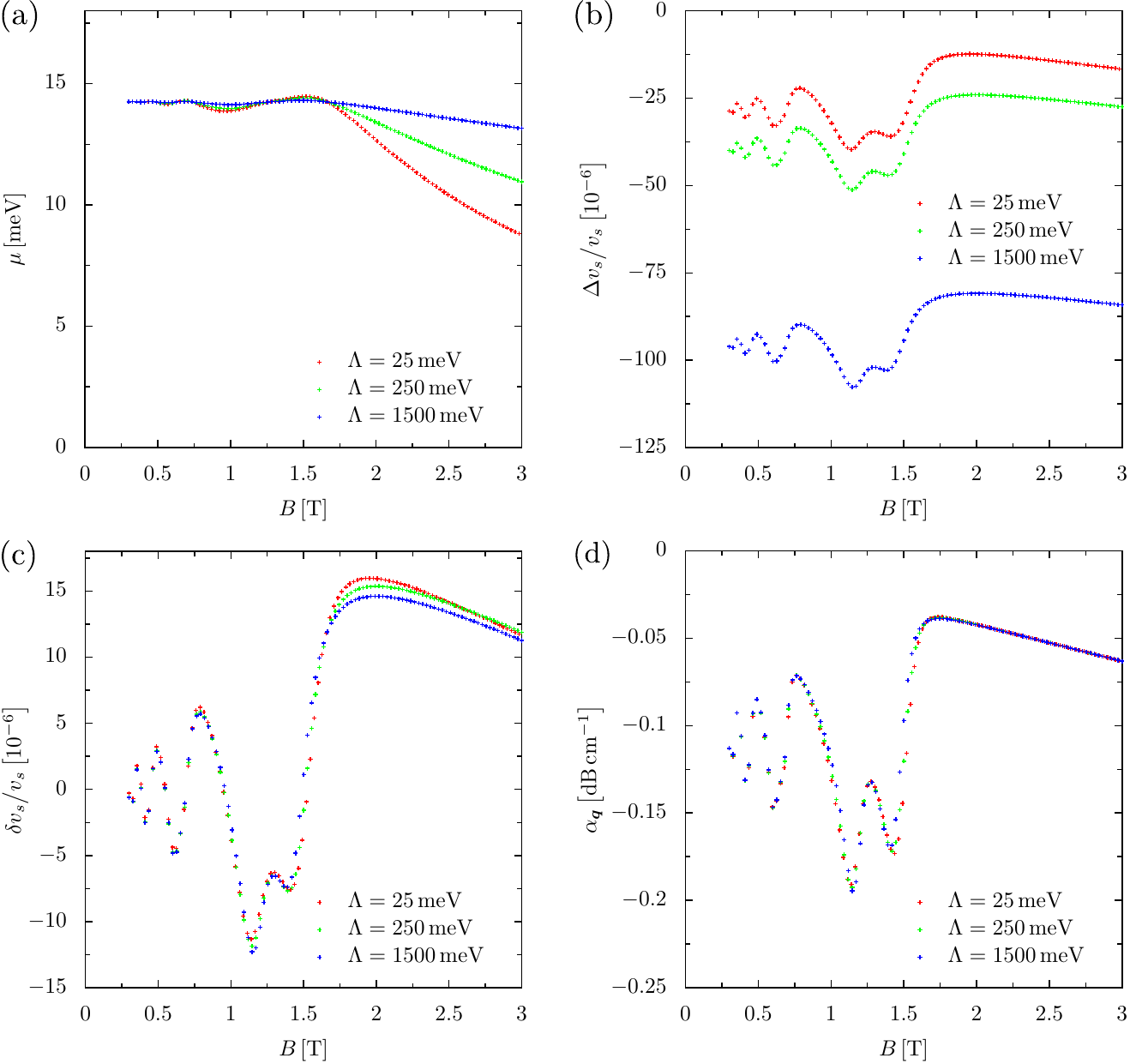}
	\caption{(a) Chemical potential $\mu(B)$, (b) absolute and (c) relative phonon-velocity renormalization $\Delta v_s(B)$ and $\delta v_s(B) = \Delta v_s(B) - \Delta v_s(B_\text{ref})$, respectively, and (d) absolute sound attenuation $\alpha_{\boldsymbol{q}}(B)$ as a function of the magnetic field for different values of the high-energy cut-off $\Lambda$ at $T=0$. The particle density has been chosen such that $\mu(B=0) = \SI{14.246}{\milli\electronvolt}$. Other parameters are as in Figs. 2, 3, and 4.}
	\label{fig:cutoff-dep}
\end{figure}

Here, we explain why the ultrasound properties are effectively independent of the high-energy cut-off $\Lambda$. This is not obvious from the outset: the phonon-velocity renormalization from Eq.~\eqref{eq:dv_finite_temp} depends explicitly on the high-energy cut-off, the sound attenuation depends implicitly on $\Lambda$ via the chemical potential. As we show in Fig.~\ref{fig:cutoff-dep}(a), however, the chemical potential can be made essentially cut-off independent up to the quantum limit if one adjusts the charge carrier density $n_0$ such that $\mu(B=0)$ is kept at a fixed value (all other system parameters remain unchanged). Using this in combination with Eq. \eqref{eq:dv_finite_temp} to calculate the phonon-velocity renormalization $\Delta v_s (B) = \Delta v_s(B_\text{ref}) + \delta v_s(B)$, we observe numerically (see Fig.~\ref{fig:cutoff-dep}(b)) that the cut-off dependence only enters the offset $\Delta v_s(B_\text{ref})$, around which $\Delta v_s(B)$ oscillates at small fields compared to the quantum-limit field, and not the magnetic-field-dependent part $\delta v_s(B)$. The oscillating term, which, as explained in the main text, is the only quantity that follows from experiments, is thus essentially cut-off independent, which is shown in Fig.~\ref{fig:cutoff-dep}(c). For the sound attenuation, the almost perfect cut-off independence holds even on the level of the absolute values $\alpha_{\boldsymbol{q}}(B)$, as we show in Fig.~\ref{fig:cutoff-dep}(d).

\section{Electron-phonon-coupling matrix elements} \label{app:el_ph_me}
Here, we calculate the electron-phonon-coupling matrix elements from Eq.~\eqref{eq:coupl_elements}. Assuming a local deformation potential and a trivial action on the orbital and spin degree of freedom, these become
\begin{equation}
	\gamma_{\nu\nu^\prime}(\boldsymbol{q})=i eD\sqrt{\frac{\hbar q}{2\rho V v_s}}\braket{\nu|e^{i\boldsymbol{q\hat{r}}}|\nu^\prime}\,,
\end{equation}
with the form factor
\begin{widetext}
	\begin{align}\label{eq:form_factor}
		\begin{split}
			\braket{k_xk_znst|e^{i\boldsymbol{q\hat{r}}}|k_x^\prime k_z^\prime n^\prime s^\prime t^\prime} = \delta_{k_x, k_x^\prime + q_x}\delta_{k_z, k_z^\prime + q_z}\,
			&\left[w^{+\uparrow}_{nst}(k_z)w^{+\uparrow}_{n^\prime s^\prime t^\prime}(k_z^\prime)M_{nn^\prime}(q_x, q_y)\right.\\
			&+w^{+\downarrow}_{nst}(k_z)w^{+\downarrow}_{n^\prime s^\prime t^\prime}(k_z^\prime)M_{n-1,n^\prime-1}(q_x, q_y)\\
			&+w^{-\uparrow}_{nst}(k_z)w^{-\uparrow}_{n^\prime s^\prime t^\prime}(k_z^\prime)M_{n-1,n^\prime-1}(q_x, q_y)\\
			&\left.+w^{-\downarrow}_{nst}(k_z)w^{-\downarrow}_{n^\prime s^\prime t^\prime}(k_z^\prime)M_{nn^\prime}(q_x, q_y)\vphantom{w^1_{nst}(k_z)}\right]\,,
		\end{split}
	\end{align}
\end{widetext}	
and where
\begin{align}
	\begin{split}
		M_{nn^\prime}(q_x, q_y) &= 
		\begin{aligned}[t]
			\int_{-\infty}^{\infty}&\mathrm{d}y \left[\vphantom{e^{iq_yy}}\right. 	e^{iq_yy}\varphi_n(y-y_0(k_x))\\
												&\times\varphi_{n^\prime}(y-y_0(k_x - q_x))\left.\vphantom{e^{iq_yy}}\right]
		\end{aligned}\\
		&=
		\begin{aligned}[t]
			&(-1)^{\max\left\{0,n-n^\prime\right\}}e^{i\phi(n^\prime-n)}\sqrt{\frac{n_{\text{min}}!}{n_{\text{max}}!}}\\
			&\times\left(\frac{|Q|}{\sqrt{2}}\right)^{|n^\prime-n|} L^{|n-n^\prime|}_{n_{\text{min}}}\left(\frac{|Q|^2}{2}\right)\,,
		\end{aligned}
	\end{split}
\end{align}
with $n_{\text{min}} = \min\left\{n,n^\prime\right\}$ and $n_{\text{max}}=\max\left\{n,n^\prime\right\}$ being the minimal and maximal Landau-band index, respectively, the norm square of the effective phonon momentum 
\begin{equation}
	|Q|^2=\ell_B^2\left(\frac{|v_x|}{v_y}q_x^2+\frac{v_y}{|v_x|}q_y^2\right)\,,
\end{equation}
and the corresponding polar angle $\tan \phi = v_y q_y/(|v_x|q_x)\,.$
We observe that the form factor from Eq.~\eqref{eq:form_factor} implements the conservation of the momentum components in $x$ and $z$ direction. The leading-order contribution to the long-wavelength expansion is given in Eq.~\eqref{eq:abs_sq_m_el}.

\onecolumngrid

\twocolumngrid

\clearpage
\newpage

\bibliography{references}

\begin{thebibliography}{50}%
\makeatletter
\providecommand \@ifxundefined [1]{%
 \@ifx{#1\undefined}
}%
\providecommand \@ifnum [1]{%
 \ifnum #1\expandafter \@firstoftwo
 \else \expandafter \@secondoftwo
 \fi
}%
\providecommand \@ifx [1]{%
 \ifx #1\expandafter \@firstoftwo
 \else \expandafter \@secondoftwo
 \fi
}%
\providecommand \natexlab [1]{#1}%
\providecommand \enquote  [1]{``#1''}%
\providecommand \bibnamefont  [1]{#1}%
\providecommand \bibfnamefont [1]{#1}%
\providecommand \citenamefont [1]{#1}%
\providecommand \href@noop [0]{\@secondoftwo}%
\providecommand \href [0]{\begingroup \@sanitize@url \@href}%
\providecommand \@href[1]{\@@startlink{#1}\@@href}%
\providecommand \@@href[1]{\endgroup#1\@@endlink}%
\providecommand \@sanitize@url [0]{\catcode `\\12\catcode `\$12\catcode
  `\&12\catcode `\#12\catcode `\^12\catcode `\_12\catcode `\%12\relax}%
\providecommand \@@startlink[1]{}%
\providecommand \@@endlink[0]{}%
\providecommand \url  [0]{\begingroup\@sanitize@url \@url }%
\providecommand \@url [1]{\endgroup\@href {#1}{\urlprefix }}%
\providecommand \urlprefix  [0]{URL }%
\providecommand \Eprint [0]{\href }%
\providecommand \doibase [0]{https://doi.org/}%
\providecommand \selectlanguage [0]{\@gobble}%
\providecommand \bibinfo  [0]{\@secondoftwo}%
\providecommand \bibfield  [0]{\@secondoftwo}%
\providecommand \translation [1]{[#1]}%
\providecommand \BibitemOpen [0]{}%
\providecommand \bibitemStop [0]{}%
\providecommand \bibitemNoStop [0]{.\EOS\space}%
\providecommand \EOS [0]{\spacefactor3000\relax}%
\providecommand \BibitemShut  [1]{\csname bibitem#1\endcsname}%
\let\auto@bib@innerbib\@empty
\bibitem [{\citenamefont {Li}\ \emph {et~al.}(2016)\citenamefont {Li},
  \citenamefont {Kharzeev}, \citenamefont {Zhang}, \citenamefont {Huang},
  \citenamefont {Pletikosi{\'{c}}}, \citenamefont {Fedorov}, \citenamefont
  {Zhong}, \citenamefont {Schneeloch}, \citenamefont {Gu},\ and\ \citenamefont
  {Valla}}]{li_chiral_2016}%
  \BibitemOpen
  \bibfield  {author} {\bibinfo {author} {\bibfnamefont {Q.}~\bibnamefont
  {Li}}, \bibinfo {author} {\bibfnamefont {D.~E.}\ \bibnamefont {Kharzeev}},
  \bibinfo {author} {\bibfnamefont {C.}~\bibnamefont {Zhang}}, \bibinfo
  {author} {\bibfnamefont {Y.}~\bibnamefont {Huang}}, \bibinfo {author}
  {\bibfnamefont {I.}~\bibnamefont {Pletikosi{\'{c}}}}, \bibinfo {author}
  {\bibfnamefont {A.~V.}\ \bibnamefont {Fedorov}}, \bibinfo {author}
  {\bibfnamefont {R.~D.}\ \bibnamefont {Zhong}}, \bibinfo {author}
  {\bibfnamefont {J.~A.}\ \bibnamefont {Schneeloch}}, \bibinfo {author}
  {\bibfnamefont {G.~D.}\ \bibnamefont {Gu}},\ and\ \bibinfo {author}
  {\bibfnamefont {T.}~\bibnamefont {Valla}},\ }\bibfield  {title} {\bibinfo
  {title} {{Chiral magnetic effect in ZrTe5}},\ }\href
  {https://doi.org/10.1038/nphys3648} {\bibfield  {journal} {\bibinfo
  {journal} {Nature Physics}\ }\textbf {\bibinfo {volume} {12}},\ \bibinfo
  {pages} {550} (\bibinfo {year} {2016})}\BibitemShut {NoStop}%
\bibitem [{\citenamefont {Chen}\ \emph {et~al.}(2015)\citenamefont {Chen},
  \citenamefont {Chen}, \citenamefont {Song}, \citenamefont {Schneeloch},
  \citenamefont {Gu}, \citenamefont {Wang},\ and\ \citenamefont
  {Wang}}]{chen_magnetoinfrared_spec_2015}%
  \BibitemOpen
  \bibfield  {author} {\bibinfo {author} {\bibfnamefont {R.~Y.}\ \bibnamefont
  {Chen}}, \bibinfo {author} {\bibfnamefont {Z.~G.}\ \bibnamefont {Chen}},
  \bibinfo {author} {\bibfnamefont {X.-Y.}\ \bibnamefont {Song}}, \bibinfo
  {author} {\bibfnamefont {J.~A.}\ \bibnamefont {Schneeloch}}, \bibinfo
  {author} {\bibfnamefont {G.~D.}\ \bibnamefont {Gu}}, \bibinfo {author}
  {\bibfnamefont {F.}~\bibnamefont {Wang}},\ and\ \bibinfo {author}
  {\bibfnamefont {N.~L.}\ \bibnamefont {Wang}},\ }\bibfield  {title} {\bibinfo
  {title} {{Magnetoinfrared Spectroscopy of Landau Levels and Zeeman Splitting
  of Three-Dimensional Massless Dirac Fermions in ${\mathrm{ZrTe}}_{5}$}},\
  }\href {https://doi.org/10.1103/PhysRevLett.115.176404} {\bibfield  {journal}
  {\bibinfo  {journal} {Phys. Rev. Lett.}\ }\textbf {\bibinfo {volume} {115}},\
  \bibinfo {pages} {176404} (\bibinfo {year} {2015})}\BibitemShut {NoStop}%
\bibitem [{\citenamefont {Zhang}\ \emph
  {et~al.}(2019{\natexlab{a}})\citenamefont {Zhang}, \citenamefont {Wang},
  \citenamefont {Guo}, \citenamefont {Zhu}, \citenamefont {Zhang},
  \citenamefont {Yang}, \citenamefont {Wang}, \citenamefont {Qu}, \citenamefont
  {Pi}, \citenamefont {Lu},\ and\ \citenamefont {Tian}}]{zhang_anomalous_2019}%
  \BibitemOpen
  \bibfield  {author} {\bibinfo {author} {\bibfnamefont {J.~L.}\ \bibnamefont
  {Zhang}}, \bibinfo {author} {\bibfnamefont {C.~M.}\ \bibnamefont {Wang}},
  \bibinfo {author} {\bibfnamefont {C.~Y.}\ \bibnamefont {Guo}}, \bibinfo
  {author} {\bibfnamefont {X.~D.}\ \bibnamefont {Zhu}}, \bibinfo {author}
  {\bibfnamefont {Y.}~\bibnamefont {Zhang}}, \bibinfo {author} {\bibfnamefont
  {J.~Y.}\ \bibnamefont {Yang}}, \bibinfo {author} {\bibfnamefont {Y.~Q.}\
  \bibnamefont {Wang}}, \bibinfo {author} {\bibfnamefont {Z.}~\bibnamefont
  {Qu}}, \bibinfo {author} {\bibfnamefont {L.}~\bibnamefont {Pi}}, \bibinfo
  {author} {\bibfnamefont {H.-Z.}\ \bibnamefont {Lu}},\ and\ \bibinfo {author}
  {\bibfnamefont {M.~L.}\ \bibnamefont {Tian}},\ }\bibfield  {title} {\bibinfo
  {title} {{Anomalous Thermoelectric Effects of ${\mathrm{ZrTe}}_{5}$ in and
  beyond the Quantum Limit}},\ }\href
  {https://doi.org/10.1103/PhysRevLett.123.196602} {\bibfield  {journal}
  {\bibinfo  {journal} {Phys. Rev. Lett.}\ }\textbf {\bibinfo {volume} {123}},\
  \bibinfo {pages} {196602} (\bibinfo {year} {2019}{\natexlab{a}})}\BibitemShut
  {NoStop}%
\bibitem [{\citenamefont {Wang}\ \emph {et~al.}(2018)\citenamefont {Wang},
  \citenamefont {Wang}, \citenamefont {Chen}, \citenamefont {Yao},\ and\
  \citenamefont {Zhou}}]{wang_first_2018}%
  \BibitemOpen
  \bibfield  {author} {\bibinfo {author} {\bibfnamefont {C.}~\bibnamefont
  {Wang}}, \bibinfo {author} {\bibfnamefont {H.}~\bibnamefont {Wang}}, \bibinfo
  {author} {\bibfnamefont {Y.~B.}\ \bibnamefont {Chen}}, \bibinfo {author}
  {\bibfnamefont {S.-H.}\ \bibnamefont {Yao}},\ and\ \bibinfo {author}
  {\bibfnamefont {J.}~\bibnamefont {Zhou}},\ }\bibfield  {title} {\bibinfo
  {title} {{First-principles study of lattice thermal conductivity in ZrTe5 and
  HfTe5}},\ }\href {https://doi.org/10.1063/1.5020615} {\bibfield  {journal}
  {\bibinfo  {journal} {Journal of Applied Physics}\ }\textbf {\bibinfo
  {volume} {123}},\ \bibinfo {pages} {175104} (\bibinfo {year}
  {2018})}\BibitemShut {NoStop}%
\bibitem [{\citenamefont {Jones}\ \emph {et~al.}(1982)\citenamefont {Jones},
  \citenamefont {Fuller}, \citenamefont {Wieting},\ and\ \citenamefont
  {Levy}}]{jones_thermoelectric_1982}%
  \BibitemOpen
  \bibfield  {author} {\bibinfo {author} {\bibfnamefont {T.}~\bibnamefont
  {Jones}}, \bibinfo {author} {\bibfnamefont {W.}~\bibnamefont {Fuller}},
  \bibinfo {author} {\bibfnamefont {T.}~\bibnamefont {Wieting}},\ and\ \bibinfo
  {author} {\bibfnamefont {F.}~\bibnamefont {Levy}},\ }\bibfield  {title}
  {\bibinfo {title} {{Thermoelectric power of HfTe5 and ZrTe5}},\ }\href
  {https://doi.org/https://doi.org/10.1016/0038-1098(82)90008-4} {\bibfield
  {journal} {\bibinfo  {journal} {Solid State Communications}\ }\textbf
  {\bibinfo {volume} {42}},\ \bibinfo {pages} {793} (\bibinfo {year}
  {1982})}\BibitemShut {NoStop}%
\bibitem [{\citenamefont {Weng}\ \emph {et~al.}(2014)\citenamefont {Weng},
  \citenamefont {Dai},\ and\ \citenamefont
  {Fang}}]{weng_transition-metal_2014}%
  \BibitemOpen
  \bibfield  {author} {\bibinfo {author} {\bibfnamefont {H.}~\bibnamefont
  {Weng}}, \bibinfo {author} {\bibfnamefont {X.}~\bibnamefont {Dai}},\ and\
  \bibinfo {author} {\bibfnamefont {Z.}~\bibnamefont {Fang}},\ }\bibfield
  {title} {\bibinfo {title} {{Transition-{Metal} {Pentatelluride}
  {$\mathrm{ZrTe}_{5}$} and {$\mathrm{HfTe}_{5}$}: {A} {Paradigm} for
  {Large}-{Gap} {Quantum} {Spin} {Hall} {Insulators}}},\ }\href
  {https://doi.org/10.1103/PhysRevX.4.011002} {\bibfield  {journal} {\bibinfo
  {journal} {Physical Review X}\ }\textbf {\bibinfo {volume} {4}},\ \bibinfo
  {pages} {011002} (\bibinfo {year} {2014})}\BibitemShut {NoStop}%
\bibitem [{\citenamefont {Zhang}\ \emph {et~al.}(2017)\citenamefont {Zhang},
  \citenamefont {Wang}, \citenamefont {Yu}, \citenamefont {Liu}, \citenamefont
  {Liang}, \citenamefont {Huang}, \citenamefont {Nie}, \citenamefont {Sun},
  \citenamefont {Zhang}, \citenamefont {Shen}, \citenamefont {Liu},
  \citenamefont {Weng}, \citenamefont {Zhao}, \citenamefont {Chen},
  \citenamefont {Jia}, \citenamefont {Hu}, \citenamefont {Ding}, \citenamefont
  {Zhao}, \citenamefont {Gao}, \citenamefont {Li}, \citenamefont {He},
  \citenamefont {Zhao}, \citenamefont {Zhang}, \citenamefont {Zhang},
  \citenamefont {Yang}, \citenamefont {Wang}, \citenamefont {Peng},
  \citenamefont {Dai}, \citenamefont {Fang}, \citenamefont {Xu}, \citenamefont
  {Chen},\ and\ \citenamefont {Zhou}}]{zhang_electronic_2017}%
  \BibitemOpen
  \bibfield  {author} {\bibinfo {author} {\bibfnamefont {Y.}~\bibnamefont
  {Zhang}}, \bibinfo {author} {\bibfnamefont {C.}~\bibnamefont {Wang}},
  \bibinfo {author} {\bibfnamefont {L.}~\bibnamefont {Yu}}, \bibinfo {author}
  {\bibfnamefont {G.}~\bibnamefont {Liu}}, \bibinfo {author} {\bibfnamefont
  {A.}~\bibnamefont {Liang}}, \bibinfo {author} {\bibfnamefont
  {J.}~\bibnamefont {Huang}}, \bibinfo {author} {\bibfnamefont
  {S.}~\bibnamefont {Nie}}, \bibinfo {author} {\bibfnamefont {X.}~\bibnamefont
  {Sun}}, \bibinfo {author} {\bibfnamefont {Y.}~\bibnamefont {Zhang}}, \bibinfo
  {author} {\bibfnamefont {B.}~\bibnamefont {Shen}}, \bibinfo {author}
  {\bibfnamefont {J.}~\bibnamefont {Liu}}, \bibinfo {author} {\bibfnamefont
  {H.}~\bibnamefont {Weng}}, \bibinfo {author} {\bibfnamefont {L.}~\bibnamefont
  {Zhao}}, \bibinfo {author} {\bibfnamefont {G.}~\bibnamefont {Chen}}, \bibinfo
  {author} {\bibfnamefont {X.}~\bibnamefont {Jia}}, \bibinfo {author}
  {\bibfnamefont {C.}~\bibnamefont {Hu}}, \bibinfo {author} {\bibfnamefont
  {Y.}~\bibnamefont {Ding}}, \bibinfo {author} {\bibfnamefont {W.}~\bibnamefont
  {Zhao}}, \bibinfo {author} {\bibfnamefont {Q.}~\bibnamefont {Gao}}, \bibinfo
  {author} {\bibfnamefont {C.}~\bibnamefont {Li}}, \bibinfo {author}
  {\bibfnamefont {S.}~\bibnamefont {He}}, \bibinfo {author} {\bibfnamefont
  {L.}~\bibnamefont {Zhao}}, \bibinfo {author} {\bibfnamefont {F.}~\bibnamefont
  {Zhang}}, \bibinfo {author} {\bibfnamefont {S.}~\bibnamefont {Zhang}},
  \bibinfo {author} {\bibfnamefont {F.}~\bibnamefont {Yang}}, \bibinfo {author}
  {\bibfnamefont {Z.}~\bibnamefont {Wang}}, \bibinfo {author} {\bibfnamefont
  {Q.}~\bibnamefont {Peng}}, \bibinfo {author} {\bibfnamefont {X.}~\bibnamefont
  {Dai}}, \bibinfo {author} {\bibfnamefont {Z.}~\bibnamefont {Fang}}, \bibinfo
  {author} {\bibfnamefont {Z.}~\bibnamefont {Xu}}, \bibinfo {author}
  {\bibfnamefont {C.}~\bibnamefont {Chen}},\ and\ \bibinfo {author}
  {\bibfnamefont {X.~J.}\ \bibnamefont {Zhou}},\ }\bibfield  {title} {\bibinfo
  {title} {{Electronic evidence of temperature-induced Lifshitz transition and
  topological nature in ZrTe5}},\ }\href {https://doi.org/10.1038/ncomms15512}
  {\bibfield  {journal} {\bibinfo  {journal} {Nature Communications}\ }\textbf
  {\bibinfo {volume} {8}},\ \bibinfo {pages} {15512} (\bibinfo {year}
  {2017})}\BibitemShut {NoStop}%
\bibitem [{\citenamefont {Xu}\ \emph {et~al.}(2018)\citenamefont {Xu},
  \citenamefont {Zhao}, \citenamefont {Marsik}, \citenamefont {Sheveleva},
  \citenamefont {Lyzwa}, \citenamefont {Dai}, \citenamefont {Chen},
  \citenamefont {Qiu},\ and\ \citenamefont {Bernhard}}]{xu_temperature_2018}%
  \BibitemOpen
  \bibfield  {author} {\bibinfo {author} {\bibfnamefont {B.}~\bibnamefont
  {Xu}}, \bibinfo {author} {\bibfnamefont {L.~X.}\ \bibnamefont {Zhao}},
  \bibinfo {author} {\bibfnamefont {P.}~\bibnamefont {Marsik}}, \bibinfo
  {author} {\bibfnamefont {E.}~\bibnamefont {Sheveleva}}, \bibinfo {author}
  {\bibfnamefont {F.}~\bibnamefont {Lyzwa}}, \bibinfo {author} {\bibfnamefont
  {Y.~M.}\ \bibnamefont {Dai}}, \bibinfo {author} {\bibfnamefont {G.~F.}\
  \bibnamefont {Chen}}, \bibinfo {author} {\bibfnamefont {X.~G.}\ \bibnamefont
  {Qiu}},\ and\ \bibinfo {author} {\bibfnamefont {C.}~\bibnamefont
  {Bernhard}},\ }\bibfield  {title} {\bibinfo {title} {{Temperature-Driven
  Topological Phase Transition and Intermediate Dirac Semimetal Phase in
  ${\mathrm{ZrTe}}_{5}$}},\ }\href
  {https://doi.org/10.1103/PhysRevLett.121.187401} {\bibfield  {journal}
  {\bibinfo  {journal} {Phys. Rev. Lett.}\ }\textbf {\bibinfo {volume} {121}},\
  \bibinfo {pages} {187401} (\bibinfo {year} {2018})}\BibitemShut {NoStop}%
\bibitem [{\citenamefont {Liu}\ \emph {et~al.}(2016)\citenamefont {Liu},
  \citenamefont {Yuan}, \citenamefont {Zhang}, \citenamefont {Jin},
  \citenamefont {Narayan}, \citenamefont {Luo}, \citenamefont {Chen},
  \citenamefont {Yang}, \citenamefont {Zou}, \citenamefont {Wu}, \citenamefont
  {Sanvito}, \citenamefont {Xia}, \citenamefont {Li}, \citenamefont {Wang},\
  and\ \citenamefont {Xiu}}]{liu_zeeman_2016}%
  \BibitemOpen
  \bibfield  {author} {\bibinfo {author} {\bibfnamefont {Y.}~\bibnamefont
  {Liu}}, \bibinfo {author} {\bibfnamefont {X.}~\bibnamefont {Yuan}}, \bibinfo
  {author} {\bibfnamefont {C.}~\bibnamefont {Zhang}}, \bibinfo {author}
  {\bibfnamefont {Z.}~\bibnamefont {Jin}}, \bibinfo {author} {\bibfnamefont
  {A.}~\bibnamefont {Narayan}}, \bibinfo {author} {\bibfnamefont
  {C.}~\bibnamefont {Luo}}, \bibinfo {author} {\bibfnamefont {Z.}~\bibnamefont
  {Chen}}, \bibinfo {author} {\bibfnamefont {L.}~\bibnamefont {Yang}}, \bibinfo
  {author} {\bibfnamefont {J.}~\bibnamefont {Zou}}, \bibinfo {author}
  {\bibfnamefont {X.}~\bibnamefont {Wu}}, \bibinfo {author} {\bibfnamefont
  {S.}~\bibnamefont {Sanvito}}, \bibinfo {author} {\bibfnamefont
  {Z.}~\bibnamefont {Xia}}, \bibinfo {author} {\bibfnamefont {L.}~\bibnamefont
  {Li}}, \bibinfo {author} {\bibfnamefont {Z.}~\bibnamefont {Wang}},\ and\
  \bibinfo {author} {\bibfnamefont {F.}~\bibnamefont {Xiu}},\ }\bibfield
  {title} {\bibinfo {title} {{Zeeman splitting and dynamical mass generation in
  Dirac semimetal ZrTe5}},\ }\href {https://doi.org/10.1038/ncomms12516}
  {\bibfield  {journal} {\bibinfo  {journal} {Nature Communications}\ }\textbf
  {\bibinfo {volume} {7}},\ \bibinfo {pages} {12516} (\bibinfo {year}
  {2016})}\BibitemShut {NoStop}%
\bibitem [{\citenamefont {Wu}\ \emph {et~al.}(2019)\citenamefont {Wu},
  \citenamefont {Zhang}, \citenamefont {Zhu}, \citenamefont {Lu}, \citenamefont
  {Zheng}, \citenamefont {Gao}, \citenamefont {Han}, \citenamefont {Zhou},
  \citenamefont {Ning},\ and\ \citenamefont {Tian}}]{wu_contactless_2019}%
  \BibitemOpen
  \bibfield  {author} {\bibinfo {author} {\bibfnamefont {M.}~\bibnamefont
  {Wu}}, \bibinfo {author} {\bibfnamefont {H.}~\bibnamefont {Zhang}}, \bibinfo
  {author} {\bibfnamefont {X.}~\bibnamefont {Zhu}}, \bibinfo {author}
  {\bibfnamefont {J.}~\bibnamefont {Lu}}, \bibinfo {author} {\bibfnamefont
  {G.}~\bibnamefont {Zheng}}, \bibinfo {author} {\bibfnamefont
  {W.}~\bibnamefont {Gao}}, \bibinfo {author} {\bibfnamefont {Y.}~\bibnamefont
  {Han}}, \bibinfo {author} {\bibfnamefont {J.}~\bibnamefont {Zhou}}, \bibinfo
  {author} {\bibfnamefont {W.}~\bibnamefont {Ning}},\ and\ \bibinfo {author}
  {\bibfnamefont {M.}~\bibnamefont {Tian}},\ }\bibfield  {title} {\bibinfo
  {title} {{Contactless Microwave Detection of Shubnikov{\textendash}De Haas
  Oscillations in Three-Dimensional Dirac Semimetal {ZrTe}5}},\ }\href
  {https://doi.org/10.1088/0256-307x/36/6/067201} {\bibfield  {journal}
  {\bibinfo  {journal} {Chinese Physics Letters}\ }\textbf {\bibinfo {volume}
  {36}},\ \bibinfo {pages} {067201} (\bibinfo {year} {2019})}\BibitemShut
  {NoStop}%
\bibitem [{\citenamefont {Liang}\ \emph {et~al.}(2018)\citenamefont {Liang},
  \citenamefont {Lin}, \citenamefont {Gibson}, \citenamefont {Kushwaha},
  \citenamefont {Liu}, \citenamefont {Wang}, \citenamefont {Xiong},
  \citenamefont {Sobota}, \citenamefont {Hashimoto}, \citenamefont {Kirchmann},
  \citenamefont {Shen}, \citenamefont {Cava},\ and\ \citenamefont
  {Ong}}]{liang_anomalous_2018}%
  \BibitemOpen
  \bibfield  {author} {\bibinfo {author} {\bibfnamefont {T.}~\bibnamefont
  {Liang}}, \bibinfo {author} {\bibfnamefont {J.}~\bibnamefont {Lin}}, \bibinfo
  {author} {\bibfnamefont {Q.}~\bibnamefont {Gibson}}, \bibinfo {author}
  {\bibfnamefont {S.}~\bibnamefont {Kushwaha}}, \bibinfo {author}
  {\bibfnamefont {M.}~\bibnamefont {Liu}}, \bibinfo {author} {\bibfnamefont
  {W.}~\bibnamefont {Wang}}, \bibinfo {author} {\bibfnamefont {H.}~\bibnamefont
  {Xiong}}, \bibinfo {author} {\bibfnamefont {J.~A.}\ \bibnamefont {Sobota}},
  \bibinfo {author} {\bibfnamefont {M.}~\bibnamefont {Hashimoto}}, \bibinfo
  {author} {\bibfnamefont {P.~S.}\ \bibnamefont {Kirchmann}}, \bibinfo {author}
  {\bibfnamefont {Z.-X.}\ \bibnamefont {Shen}}, \bibinfo {author}
  {\bibfnamefont {R.~J.}\ \bibnamefont {Cava}},\ and\ \bibinfo {author}
  {\bibfnamefont {N.~P.}\ \bibnamefont {Ong}},\ }\bibfield  {title} {\bibinfo
  {title} {{Anomalous Hall effect in ZrTe5}},\ }\href
  {https://doi.org/10.1038/s41567-018-0078-z} {\bibfield  {journal} {\bibinfo
  {journal} {Nature Physics}\ }\textbf {\bibinfo {volume} {14}},\ \bibinfo
  {pages} {451} (\bibinfo {year} {2018})}\BibitemShut {NoStop}%
\bibitem [{\citenamefont {Tang}\ \emph {et~al.}(2019)\citenamefont {Tang},
  \citenamefont {Ren}, \citenamefont {Wang}, \citenamefont {Zhong},
  \citenamefont {Schneeloch}, \citenamefont {Yang}, \citenamefont {Yang},
  \citenamefont {Lee}, \citenamefont {Gu}, \citenamefont {Qiao},\ and\
  \citenamefont {Zhang}}]{tang_three_dimensional_2019}%
  \BibitemOpen
  \bibfield  {author} {\bibinfo {author} {\bibfnamefont {F.}~\bibnamefont
  {Tang}}, \bibinfo {author} {\bibfnamefont {Y.}~\bibnamefont {Ren}}, \bibinfo
  {author} {\bibfnamefont {P.}~\bibnamefont {Wang}}, \bibinfo {author}
  {\bibfnamefont {R.}~\bibnamefont {Zhong}}, \bibinfo {author} {\bibfnamefont
  {J.}~\bibnamefont {Schneeloch}}, \bibinfo {author} {\bibfnamefont {S.~A.}\
  \bibnamefont {Yang}}, \bibinfo {author} {\bibfnamefont {K.}~\bibnamefont
  {Yang}}, \bibinfo {author} {\bibfnamefont {P.~A.}\ \bibnamefont {Lee}},
  \bibinfo {author} {\bibfnamefont {G.}~\bibnamefont {Gu}}, \bibinfo {author}
  {\bibfnamefont {Z.}~\bibnamefont {Qiao}},\ and\ \bibinfo {author}
  {\bibfnamefont {L.}~\bibnamefont {Zhang}},\ }\bibfield  {title} {\bibinfo
  {title} {{Three-dimensional quantum Hall effect and metal--insulator
  transition in ZrTe5}},\ }\href {https://doi.org/10.1038/s41586-019-1180-9}
  {\bibfield  {journal} {\bibinfo  {journal} {Nature}\ }\textbf {\bibinfo
  {volume} {569}},\ \bibinfo {pages} {537} (\bibinfo {year}
  {2019})}\BibitemShut {NoStop}%
\bibitem [{\citenamefont {Galeski}\ \emph {et~al.}(2021)\citenamefont
  {Galeski}, \citenamefont {Ehmcke}, \citenamefont {Wawrzyńczak},
  \citenamefont {Lozano}, \citenamefont {Cho}, \citenamefont {Sharma},
  \citenamefont {Das}, \citenamefont {Küster}, \citenamefont {Sessi},
  \citenamefont {Brando}, \citenamefont {Küchler}, \citenamefont {Markou},
  \citenamefont {König}, \citenamefont {Swekis}, \citenamefont {Felser},
  \citenamefont {Sassa}, \citenamefont {Li}, \citenamefont {Gu}, \citenamefont
  {Zimmermann}, \citenamefont {Ivashko}, \citenamefont {Gorbunov},
  \citenamefont {Zherlitsyn}, \citenamefont {Förster}, \citenamefont {Parkin},
  \citenamefont {Wosnitza}, \citenamefont {Meng},\ and\ \citenamefont
  {Gooth}}]{galeski_origin_2021}%
  \BibitemOpen
  \bibfield  {author} {\bibinfo {author} {\bibfnamefont {S.}~\bibnamefont
  {Galeski}}, \bibinfo {author} {\bibfnamefont {T.}~\bibnamefont {Ehmcke}},
  \bibinfo {author} {\bibfnamefont {R.}~\bibnamefont {Wawrzyńczak}}, \bibinfo
  {author} {\bibfnamefont {P.~M.}\ \bibnamefont {Lozano}}, \bibinfo {author}
  {\bibfnamefont {K.}~\bibnamefont {Cho}}, \bibinfo {author} {\bibfnamefont
  {A.}~\bibnamefont {Sharma}}, \bibinfo {author} {\bibfnamefont
  {S.}~\bibnamefont {Das}}, \bibinfo {author} {\bibfnamefont {F.}~\bibnamefont
  {Küster}}, \bibinfo {author} {\bibfnamefont {P.}~\bibnamefont {Sessi}},
  \bibinfo {author} {\bibfnamefont {M.}~\bibnamefont {Brando}}, \bibinfo
  {author} {\bibfnamefont {R.}~\bibnamefont {Küchler}}, \bibinfo {author}
  {\bibfnamefont {A.}~\bibnamefont {Markou}}, \bibinfo {author} {\bibfnamefont
  {M.}~\bibnamefont {König}}, \bibinfo {author} {\bibfnamefont
  {P.}~\bibnamefont {Swekis}}, \bibinfo {author} {\bibfnamefont
  {C.}~\bibnamefont {Felser}}, \bibinfo {author} {\bibfnamefont
  {Y.}~\bibnamefont {Sassa}}, \bibinfo {author} {\bibfnamefont
  {Q.}~\bibnamefont {Li}}, \bibinfo {author} {\bibfnamefont {G.}~\bibnamefont
  {Gu}}, \bibinfo {author} {\bibfnamefont {M.~V.}\ \bibnamefont {Zimmermann}},
  \bibinfo {author} {\bibfnamefont {O.}~\bibnamefont {Ivashko}}, \bibinfo
  {author} {\bibfnamefont {D.~I.}\ \bibnamefont {Gorbunov}}, \bibinfo {author}
  {\bibfnamefont {S.}~\bibnamefont {Zherlitsyn}}, \bibinfo {author}
  {\bibfnamefont {T.}~\bibnamefont {Förster}}, \bibinfo {author}
  {\bibfnamefont {S.~S.~P.}\ \bibnamefont {Parkin}}, \bibinfo {author}
  {\bibfnamefont {J.}~\bibnamefont {Wosnitza}}, \bibinfo {author}
  {\bibfnamefont {T.}~\bibnamefont {Meng}},\ and\ \bibinfo {author}
  {\bibfnamefont {J.}~\bibnamefont {Gooth}},\ }\bibfield  {title} {\bibinfo
  {title} {{Origin of the quasi-quantized {Hall} effect in {ZrTe5}}},\ }\href
  {https://doi.org/10.1038/s41467-021-23435-y} {\bibfield  {journal} {\bibinfo
  {journal} {Nature Communications}\ }\textbf {\bibinfo {volume} {12}},\
  \bibinfo {pages} {3197} (\bibinfo {year} {2021})}\BibitemShut {NoStop}%
\bibitem [{\citenamefont {Galeski}\ \emph {et~al.}(2020)\citenamefont
  {Galeski}, \citenamefont {Zhao}, \citenamefont {Wawrzy{\'{n}}czak},
  \citenamefont {Meng}, \citenamefont {F{\"o}rster}, \citenamefont {Lozano},
  \citenamefont {Honnali}, \citenamefont {Lamba}, \citenamefont {Ehmcke},
  \citenamefont {Markou}, \citenamefont {Li.}, \citenamefont {Gu},
  \citenamefont {Zhu}, \citenamefont {Wosnitza}, \citenamefont {Felser},
  \citenamefont {Chen},\ and\ \citenamefont
  {Gooth}}]{galeski_unconventional_2020}%
  \BibitemOpen
  \bibfield  {author} {\bibinfo {author} {\bibfnamefont {S.}~\bibnamefont
  {Galeski}}, \bibinfo {author} {\bibfnamefont {X.}~\bibnamefont {Zhao}},
  \bibinfo {author} {\bibfnamefont {R.}~\bibnamefont {Wawrzy{\'{n}}czak}},
  \bibinfo {author} {\bibfnamefont {T.}~\bibnamefont {Meng}}, \bibinfo {author}
  {\bibfnamefont {T.}~\bibnamefont {F{\"o}rster}}, \bibinfo {author}
  {\bibfnamefont {P.~M.}\ \bibnamefont {Lozano}}, \bibinfo {author}
  {\bibfnamefont {S.}~\bibnamefont {Honnali}}, \bibinfo {author} {\bibfnamefont
  {N.}~\bibnamefont {Lamba}}, \bibinfo {author} {\bibfnamefont
  {T.}~\bibnamefont {Ehmcke}}, \bibinfo {author} {\bibfnamefont
  {A.}~\bibnamefont {Markou}}, \bibinfo {author} {\bibfnamefont
  {Q.}~\bibnamefont {Li.}}, \bibinfo {author} {\bibfnamefont {G.}~\bibnamefont
  {Gu}}, \bibinfo {author} {\bibfnamefont {W.}~\bibnamefont {Zhu}}, \bibinfo
  {author} {\bibfnamefont {J.}~\bibnamefont {Wosnitza}}, \bibinfo {author}
  {\bibfnamefont {C.}~\bibnamefont {Felser}}, \bibinfo {author} {\bibfnamefont
  {G.~F.}\ \bibnamefont {Chen}},\ and\ \bibinfo {author} {\bibfnamefont
  {J.}~\bibnamefont {Gooth}},\ }\bibfield  {title} {\bibinfo {title}
  {{Unconventional Hall response in the quantum limit of HfTe5}},\ }\href
  {https://doi.org/10.1038/s41467-020-19773-y} {\bibfield  {journal} {\bibinfo
  {journal} {Nature Communications}\ }\textbf {\bibinfo {volume} {11}},\
  \bibinfo {pages} {5926} (\bibinfo {year} {2020})}\BibitemShut {NoStop}%
\bibitem [{\citenamefont {Wang}\ \emph {et~al.}(2020)\citenamefont {Wang},
  \citenamefont {Ren}, \citenamefont {Tang}, \citenamefont {Wang},
  \citenamefont {Hou}, \citenamefont {Zeng}, \citenamefont {Zhang},\ and\
  \citenamefont {Qiao}}]{wang_approaching_2020}%
  \BibitemOpen
  \bibfield  {author} {\bibinfo {author} {\bibfnamefont {P.}~\bibnamefont
  {Wang}}, \bibinfo {author} {\bibfnamefont {Y.}~\bibnamefont {Ren}}, \bibinfo
  {author} {\bibfnamefont {F.}~\bibnamefont {Tang}}, \bibinfo {author}
  {\bibfnamefont {P.}~\bibnamefont {Wang}}, \bibinfo {author} {\bibfnamefont
  {T.}~\bibnamefont {Hou}}, \bibinfo {author} {\bibfnamefont {H.}~\bibnamefont
  {Zeng}}, \bibinfo {author} {\bibfnamefont {L.}~\bibnamefont {Zhang}},\ and\
  \bibinfo {author} {\bibfnamefont {Z.}~\bibnamefont {Qiao}},\ }\bibfield
  {title} {\bibinfo {title} {{Approaching three-dimensional quantum Hall effect
  in bulk $\mathrm{HfT}{\mathrm{e}}_{5}$}},\ }\href
  {https://doi.org/10.1103/PhysRevB.101.161201} {\bibfield  {journal} {\bibinfo
   {journal} {Phys. Rev. B}\ }\textbf {\bibinfo {volume} {101}},\ \bibinfo
  {pages} {161201(R)} (\bibinfo {year} {2020})}\BibitemShut {NoStop}%
\bibitem [{\citenamefont {Okada}\ \emph {et~al.}(1982)\citenamefont {Okada},
  \citenamefont {Sambongi}, \citenamefont {Ido}, \citenamefont {Tazuke},
  \citenamefont {Aoki},\ and\ \citenamefont {Fujita}}]{okada_negative_1982}%
  \BibitemOpen
  \bibfield  {author} {\bibinfo {author} {\bibfnamefont {S.}~\bibnamefont
  {Okada}}, \bibinfo {author} {\bibfnamefont {T.}~\bibnamefont {Sambongi}},
  \bibinfo {author} {\bibfnamefont {M.}~\bibnamefont {Ido}}, \bibinfo {author}
  {\bibfnamefont {Y.}~\bibnamefont {Tazuke}}, \bibinfo {author} {\bibfnamefont
  {R.}~\bibnamefont {Aoki}},\ and\ \bibinfo {author} {\bibfnamefont
  {O.}~\bibnamefont {Fujita}},\ }\bibfield  {title} {\bibinfo {title}
  {{Negative Evidences for Charge/Spin Density Wave in ZrTe5}},\ }\href
  {https://doi.org/10.1143/JPSJ.51.460} {\bibfield  {journal} {\bibinfo
  {journal} {Journal of the Physical Society of Japan}\ }\textbf {\bibinfo
  {volume} {51}},\ \bibinfo {pages} {460} (\bibinfo {year} {1982})}\BibitemShut
  {NoStop}%
\bibitem [{\citenamefont {Trescher}\ \emph {et~al.}(2017)\citenamefont
  {Trescher}, \citenamefont {Bergholtz}, \citenamefont {Udagawa},\ and\
  \citenamefont {Knolle}}]{trescher_cdw_2017}%
  \BibitemOpen
  \bibfield  {author} {\bibinfo {author} {\bibfnamefont {M.}~\bibnamefont
  {Trescher}}, \bibinfo {author} {\bibfnamefont {E.~J.}\ \bibnamefont
  {Bergholtz}}, \bibinfo {author} {\bibfnamefont {M.}~\bibnamefont {Udagawa}},\
  and\ \bibinfo {author} {\bibfnamefont {J.}~\bibnamefont {Knolle}},\
  }\bibfield  {title} {\bibinfo {title} {{Charge density wave instabilities of
  type-II Weyl semimetals in a strong magnetic field}},\ }\href
  {https://doi.org/10.1103/PhysRevB.96.201101} {\bibfield  {journal} {\bibinfo
  {journal} {Phys. Rev. B}\ }\textbf {\bibinfo {volume} {96}},\ \bibinfo
  {pages} {201101(R)} (\bibinfo {year} {2017})}\BibitemShut {NoStop}%
\bibitem [{\citenamefont {Trescher}\ \emph {et~al.}(2018)\citenamefont
  {Trescher}, \citenamefont {Bergholtz},\ and\ \citenamefont
  {Knolle}}]{trescher_quantum_2018}%
  \BibitemOpen
  \bibfield  {author} {\bibinfo {author} {\bibfnamefont {M.}~\bibnamefont
  {Trescher}}, \bibinfo {author} {\bibfnamefont {E.~J.}\ \bibnamefont
  {Bergholtz}},\ and\ \bibinfo {author} {\bibfnamefont {J.}~\bibnamefont
  {Knolle}},\ }\bibfield  {title} {\bibinfo {title} {{Quantum oscillations and
  magnetoresistance in type-II Weyl semimetals: Effect of a field-induced
  charge density wave}},\ }\href {https://doi.org/10.1103/PhysRevB.98.125304}
  {\bibfield  {journal} {\bibinfo  {journal} {Phys. Rev. B}\ }\textbf {\bibinfo
  {volume} {98}},\ \bibinfo {pages} {125304} (\bibinfo {year}
  {2018})}\BibitemShut {NoStop}%
\bibitem [{\citenamefont {Mahan}(2010)}]{mahan_many_2010}%
  \BibitemOpen
  \bibfield  {author} {\bibinfo {author} {\bibfnamefont {G.~D.}\ \bibnamefont
  {Mahan}},\ }\href@noop {} {\emph {\bibinfo {title} {{Many-Particle
  Physics}}}}\ (\bibinfo  {publisher} {Springer US},\ \bibinfo {address}
  {USA},\ \bibinfo {year} {2010})\BibitemShut {NoStop}%
\bibitem [{\citenamefont {Jericho}\ \emph {et~al.}(1980)\citenamefont
  {Jericho}, \citenamefont {Simpson},\ and\ \citenamefont
  {Frindt}}]{jericho_velocity_1980}%
  \BibitemOpen
  \bibfield  {author} {\bibinfo {author} {\bibfnamefont {M.~H.}\ \bibnamefont
  {Jericho}}, \bibinfo {author} {\bibfnamefont {A.~M.}\ \bibnamefont
  {Simpson}},\ and\ \bibinfo {author} {\bibfnamefont {R.~F.}\ \bibnamefont
  {Frindt}},\ }\bibfield  {title} {\bibinfo {title} {{Velocity of ultrasonic
  waves in $2H$-Nb${\mathrm{Se}}_{2}$, $2H$-Ta${\mathrm{S}}_{2}$, and
  $1T$-Ta${\mathrm{S}}_{2}$}},\ }\href
  {https://doi.org/10.1103/PhysRevB.22.4907} {\bibfield  {journal} {\bibinfo
  {journal} {Phys. Rev. B}\ }\textbf {\bibinfo {volume} {22}},\ \bibinfo
  {pages} {4907} (\bibinfo {year} {1980})}\BibitemShut {NoStop}%
\bibitem [{\citenamefont {Song}\ \emph {et~al.}(2017)\citenamefont {Song},
  \citenamefont {Fang},\ and\ \citenamefont {Dai}}]{song_instability_2017}%
  \BibitemOpen
  \bibfield  {author} {\bibinfo {author} {\bibfnamefont {Z.}~\bibnamefont
  {Song}}, \bibinfo {author} {\bibfnamefont {Z.}~\bibnamefont {Fang}},\ and\
  \bibinfo {author} {\bibfnamefont {X.}~\bibnamefont {Dai}},\ }\bibfield
  {title} {\bibinfo {title} {{Instability of Dirac semimetal phase under a
  strong magnetic field}},\ }\href {https://doi.org/10.1103/PhysRevB.96.235104}
  {\bibfield  {journal} {\bibinfo  {journal} {Phys. Rev. B}\ }\textbf {\bibinfo
  {volume} {96}},\ \bibinfo {pages} {235104} (\bibinfo {year}
  {2017})}\BibitemShut {NoStop}%
\bibitem [{\citenamefont {Saint-Paul}\ \emph {et~al.}(2016)\citenamefont
  {Saint-Paul}, \citenamefont {Guttin}, \citenamefont {Lejay}, \citenamefont
  {Remenyi}, \citenamefont {Leynaud},\ and\ \citenamefont
  {Monceau}}]{saintpaul_elastic_2016}%
  \BibitemOpen
  \bibfield  {author} {\bibinfo {author} {\bibfnamefont {M.}~\bibnamefont
  {Saint-Paul}}, \bibinfo {author} {\bibfnamefont {C.}~\bibnamefont {Guttin}},
  \bibinfo {author} {\bibfnamefont {P.}~\bibnamefont {Lejay}}, \bibinfo
  {author} {\bibfnamefont {G.}~\bibnamefont {Remenyi}}, \bibinfo {author}
  {\bibfnamefont {O.}~\bibnamefont {Leynaud}},\ and\ \bibinfo {author}
  {\bibfnamefont {P.}~\bibnamefont {Monceau}},\ }\bibfield  {title} {\bibinfo
  {title} {{Elastic anomalies at the charge density wave transition in
  TbTe3}},\ }\href {https://doi.org/https://doi.org/10.1016/j.ssc.2016.02.008}
  {\bibfield  {journal} {\bibinfo  {journal} {Solid State Communications}\
  }\textbf {\bibinfo {volume} {233}},\ \bibinfo {pages} {24} (\bibinfo {year}
  {2016})}\BibitemShut {NoStop}%
\bibitem [{\citenamefont {Tsang}\ \emph {et~al.}(1976)\citenamefont {Tsang},
  \citenamefont {Smith},\ and\ \citenamefont {Shafer}}]{tsang_raman_1976}%
  \BibitemOpen
  \bibfield  {author} {\bibinfo {author} {\bibfnamefont {J.~C.}\ \bibnamefont
  {Tsang}}, \bibinfo {author} {\bibfnamefont {J.~E.}\ \bibnamefont {Smith}},\
  and\ \bibinfo {author} {\bibfnamefont {M.~W.}\ \bibnamefont {Shafer}},\
  }\bibfield  {title} {\bibinfo {title} {{Raman Spectroscopy of Soft Modes at
  the Charge-Density-Wave Phase Transition in
  $2H\ensuremath{-}\mathrm{Nb}{\mathrm{Se}}_{2}$}},\ }\href
  {https://doi.org/10.1103/PhysRevLett.37.1407} {\bibfield  {journal} {\bibinfo
   {journal} {Phys. Rev. Lett.}\ }\textbf {\bibinfo {volume} {37}},\ \bibinfo
  {pages} {1407} (\bibinfo {year} {1976})}\BibitemShut {NoStop}%
\bibitem [{\citenamefont {Qin}\ \emph {et~al.}(2020)\citenamefont {Qin},
  \citenamefont {Li}, \citenamefont {Du}, \citenamefont {Wang}, \citenamefont
  {Zhang}, \citenamefont {Yu}, \citenamefont {Lu},\ and\ \citenamefont
  {Xie}}]{Qin_cdw_2020}%
  \BibitemOpen
  \bibfield  {author} {\bibinfo {author} {\bibfnamefont {F.}~\bibnamefont
  {Qin}}, \bibinfo {author} {\bibfnamefont {S.}~\bibnamefont {Li}}, \bibinfo
  {author} {\bibfnamefont {Z.~Z.}\ \bibnamefont {Du}}, \bibinfo {author}
  {\bibfnamefont {C.~M.}\ \bibnamefont {Wang}}, \bibinfo {author}
  {\bibfnamefont {W.}~\bibnamefont {Zhang}}, \bibinfo {author} {\bibfnamefont
  {D.}~\bibnamefont {Yu}}, \bibinfo {author} {\bibfnamefont {H.-Z.}\
  \bibnamefont {Lu}},\ and\ \bibinfo {author} {\bibfnamefont {X.~C.}\
  \bibnamefont {Xie}},\ }\bibfield  {title} {\bibinfo {title} {{Theory for the
  Charge-Density-Wave Mechanism of 3D Quantum Hall Effect}},\ }\href
  {https://doi.org/10.1103/PhysRevLett.125.206601} {\bibfield  {journal}
  {\bibinfo  {journal} {Phys. Rev. Lett.}\ }\textbf {\bibinfo {volume} {125}},\
  \bibinfo {pages} {206601} (\bibinfo {year} {2020})}\BibitemShut {NoStop}%
\bibitem [{\citenamefont {Zhao}\ \emph {et~al.}(2021)\citenamefont {Zhao},
  \citenamefont {Lu},\ and\ \citenamefont {Xie}}]{Zhao.2021}%
  \BibitemOpen
  \bibfield  {author} {\bibinfo {author} {\bibfnamefont {P.-L.}\ \bibnamefont
  {Zhao}}, \bibinfo {author} {\bibfnamefont {H.-Z.}\ \bibnamefont {Lu}},\ and\
  \bibinfo {author} {\bibfnamefont {X.~C.}\ \bibnamefont {Xie}},\ }\bibfield
  {title} {\bibinfo {title} {{Theory for Magnetic-Field-Driven 3D
  Metal-Insulator Transitions in the Quantum Limit}},\ }\href
  {https://doi.org/10.1103/PhysRevLett.127.046602} {\bibfield  {journal}
  {\bibinfo  {journal} {{Physical Review Letters}}\ }\textbf {\bibinfo {volume}
  {127}},\ \bibinfo {pages} {046602} (\bibinfo {year} {2021})}\BibitemShut
  {NoStop}%
\bibitem [{\citenamefont {Tian}\ \emph {et~al.}(2021)\citenamefont {Tian},
  \citenamefont {Ghassemi},\ and\ \citenamefont
  {Ross}}]{tian_gap_opening_2021}%
  \BibitemOpen
  \bibfield  {author} {\bibinfo {author} {\bibfnamefont {Y.}~\bibnamefont
  {Tian}}, \bibinfo {author} {\bibfnamefont {N.}~\bibnamefont {Ghassemi}},\
  and\ \bibinfo {author} {\bibfnamefont {Joseph H.}\ \bibnamefont {Ross}},\
  }\bibfield  {title} {\bibinfo {title} {{Gap-Opening Transition in Dirac
  Semimetal ${\mathrm{ZrTe}}_{5}$}},\ }\href
  {https://doi.org/10.1103/PhysRevLett.126.236401} {\bibfield  {journal}
  {\bibinfo  {journal} {Phys. Rev. Lett.}\ }\textbf {\bibinfo {volume} {126}},\
  \bibinfo {pages} {236401} (\bibinfo {year} {2021})}\BibitemShut {NoStop}%
\bibitem [{\citenamefont {Shapourian}\ \emph {et~al.}(2015)\citenamefont
  {Shapourian}, \citenamefont {Hughes},\ and\ \citenamefont
  {Ryu}}]{shapourian_visco_2015}%
  \BibitemOpen
  \bibfield  {author} {\bibinfo {author} {\bibfnamefont {H.}~\bibnamefont
  {Shapourian}}, \bibinfo {author} {\bibfnamefont {T.~L.}\ \bibnamefont
  {Hughes}},\ and\ \bibinfo {author} {\bibfnamefont {S.}~\bibnamefont {Ryu}},\
  }\bibfield  {title} {\bibinfo {title} {{Viscoelastic response of topological
  tight-binding models in two and three dimensions}},\ }\href
  {https://doi.org/10.1103/PhysRevB.92.165131} {\bibfield  {journal} {\bibinfo
  {journal} {Phys. Rev. B}\ }\textbf {\bibinfo {volume} {92}},\ \bibinfo
  {pages} {165131} (\bibinfo {year} {2015})}\BibitemShut {NoStop}%
\bibitem [{\citenamefont {Rinkel}\ \emph {et~al.}(2017)\citenamefont {Rinkel},
  \citenamefont {Lopes},\ and\ \citenamefont
  {Garate}}]{rinkel_signatures_2017}%
  \BibitemOpen
  \bibfield  {author} {\bibinfo {author} {\bibfnamefont {P.}~\bibnamefont
  {Rinkel}}, \bibinfo {author} {\bibfnamefont {P.~L.~S.}\ \bibnamefont
  {Lopes}},\ and\ \bibinfo {author} {\bibfnamefont {I.}~\bibnamefont
  {Garate}},\ }\bibfield  {title} {\bibinfo {title} {{Signatures of the Chiral
  Anomaly in Phonon Dynamics}},\ }\href
  {https://doi.org/10.1103/PhysRevLett.119.107401} {\bibfield  {journal}
  {\bibinfo  {journal} {Phys. Rev. Lett.}\ }\textbf {\bibinfo {volume} {119}},\
  \bibinfo {pages} {107401} (\bibinfo {year} {2017})}\BibitemShut {NoStop}%
\bibitem [{\citenamefont {Zhang}\ and\ \citenamefont
  {Zhou}(2020)}]{zhang_quantum_2020}%
  \BibitemOpen
  \bibfield  {author} {\bibinfo {author} {\bibfnamefont {S.-B.}\ \bibnamefont
  {Zhang}}\ and\ \bibinfo {author} {\bibfnamefont {J.}~\bibnamefont {Zhou}},\
  }\bibfield  {title} {\bibinfo {title} {{Quantum oscillations in acoustic
  phonons in Weyl semimetals}},\ }\href
  {https://doi.org/10.1103/PhysRevB.101.085202} {\bibfield  {journal} {\bibinfo
   {journal} {Phys. Rev. B}\ }\textbf {\bibinfo {volume} {101}},\ \bibinfo
  {pages} {085202} (\bibinfo {year} {2020})}\BibitemShut {NoStop}%
\bibitem [{\citenamefont {Schindler}\ \emph {et~al.}(2020)\citenamefont
  {Schindler}, \citenamefont {Gorbunov}, \citenamefont {Zherlitsyn},
  \citenamefont {Galeski}, \citenamefont {Schmidt}, \citenamefont {Wosnitza},\
  and\ \citenamefont {Gooth}}]{schindler_strong_2020}%
  \BibitemOpen
  \bibfield  {author} {\bibinfo {author} {\bibfnamefont {C.}~\bibnamefont
  {Schindler}}, \bibinfo {author} {\bibfnamefont {D.}~\bibnamefont {Gorbunov}},
  \bibinfo {author} {\bibfnamefont {S.}~\bibnamefont {Zherlitsyn}}, \bibinfo
  {author} {\bibfnamefont {S.}~\bibnamefont {Galeski}}, \bibinfo {author}
  {\bibfnamefont {M.}~\bibnamefont {Schmidt}}, \bibinfo {author} {\bibfnamefont
  {J.}~\bibnamefont {Wosnitza}},\ and\ \bibinfo {author} {\bibfnamefont
  {J.}~\bibnamefont {Gooth}},\ }\bibfield  {title} {\bibinfo {title} {{Strong
  anisotropy of the electron-phonon interaction in NbP probed by
  magnetoacoustic quantum oscillations}},\ }\href
  {https://doi.org/10.1103/PhysRevB.102.165156} {\bibfield  {journal} {\bibinfo
   {journal} {Phys. Rev. B}\ }\textbf {\bibinfo {volume} {102}},\ \bibinfo
  {pages} {165156} (\bibinfo {year} {2020})}\BibitemShut {NoStop}%
\bibitem [{\citenamefont {Lalibert\'e}\ \emph {et~al.}(2020)\citenamefont
  {Lalibert\'e}, \citenamefont {B\'elanger}, \citenamefont {Nair},
  \citenamefont {Analytis}, \citenamefont {Boulanger}, \citenamefont {Dion},
  \citenamefont {Taillefer},\ and\ \citenamefont
  {Quilliam}}]{laliberte_field_2020}%
  \BibitemOpen
  \bibfield  {author} {\bibinfo {author} {\bibfnamefont {F.}~\bibnamefont
  {Lalibert\'e}}, \bibinfo {author} {\bibfnamefont {F.}~\bibnamefont
  {B\'elanger}}, \bibinfo {author} {\bibfnamefont {N.~L.}\ \bibnamefont
  {Nair}}, \bibinfo {author} {\bibfnamefont {J.~G.}\ \bibnamefont {Analytis}},
  \bibinfo {author} {\bibfnamefont {M.-E.}\ \bibnamefont {Boulanger}}, \bibinfo
  {author} {\bibfnamefont {M.}~\bibnamefont {Dion}}, \bibinfo {author}
  {\bibfnamefont {L.}~\bibnamefont {Taillefer}},\ and\ \bibinfo {author}
  {\bibfnamefont {J.~A.}\ \bibnamefont {Quilliam}},\ }\bibfield  {title}
  {\bibinfo {title} {{Field-angle dependence of sound velocity in the Weyl
  semimetal TaAs}},\ }\href {https://doi.org/10.1103/PhysRevB.102.125104}
  {\bibfield  {journal} {\bibinfo  {journal} {Phys. Rev. B}\ }\textbf {\bibinfo
  {volume} {102}},\ \bibinfo {pages} {125104} (\bibinfo {year}
  {2020})}\BibitemShut {NoStop}%
\bibitem [{\citenamefont {Rinkel}\ \emph {et~al.}(2019)\citenamefont {Rinkel},
  \citenamefont {Lopes},\ and\ \citenamefont {Garate}}]{rinkel_influence_2019}%
  \BibitemOpen
  \bibfield  {author} {\bibinfo {author} {\bibfnamefont {P.}~\bibnamefont
  {Rinkel}}, \bibinfo {author} {\bibfnamefont {P.~L.~S.}\ \bibnamefont
  {Lopes}},\ and\ \bibinfo {author} {\bibfnamefont {I.}~\bibnamefont
  {Garate}},\ }\bibfield  {title} {\bibinfo {title} {{Influence of Landau
  levels on the phonon dispersion of Weyl semimetals}},\ }\href
  {https://doi.org/10.1103/PhysRevB.99.144301} {\bibfield  {journal} {\bibinfo
  {journal} {Phys. Rev. B}\ }\textbf {\bibinfo {volume} {99}},\ \bibinfo
  {pages} {144301} (\bibinfo {year} {2019})}\BibitemShut {NoStop}%
\bibitem [{\citenamefont {Spivak}\ and\ \citenamefont
  {Andreev}(2016)}]{spivak_magneto_2016}%
  \BibitemOpen
  \bibfield  {author} {\bibinfo {author} {\bibfnamefont {B.~Z.}\ \bibnamefont
  {Spivak}}\ and\ \bibinfo {author} {\bibfnamefont {A.~V.}\ \bibnamefont
  {Andreev}},\ }\bibfield  {title} {\bibinfo {title} {{Magnetotransport
  phenomena related to the chiral anomaly in Weyl semimetals}},\ }\href
  {https://doi.org/10.1103/PhysRevB.93.085107} {\bibfield  {journal} {\bibinfo
  {journal} {Phys. Rev. B}\ }\textbf {\bibinfo {volume} {93}},\ \bibinfo
  {pages} {085107} (\bibinfo {year} {2016})}\BibitemShut {NoStop}%
\bibitem [{\citenamefont {Pikulin}\ \emph {et~al.}(2016)\citenamefont
  {Pikulin}, \citenamefont {Chen},\ and\ \citenamefont
  {Franz}}]{pikulin_chiral_2016}%
  \BibitemOpen
  \bibfield  {author} {\bibinfo {author} {\bibfnamefont {D.~I.}\ \bibnamefont
  {Pikulin}}, \bibinfo {author} {\bibfnamefont {A.}~\bibnamefont {Chen}},\ and\
  \bibinfo {author} {\bibfnamefont {M.}~\bibnamefont {Franz}},\ }\bibfield
  {title} {\bibinfo {title} {{Chiral Anomaly from Strain-Induced Gauge Fields
  in Dirac and Weyl Semimetals}},\ }\href
  {https://doi.org/10.1103/PhysRevX.6.041021} {\bibfield  {journal} {\bibinfo
  {journal} {Phys. Rev. X}\ }\textbf {\bibinfo {volume} {6}},\ \bibinfo {pages}
  {041021} (\bibinfo {year} {2016})}\BibitemShut {NoStop}%
\bibitem [{\citenamefont {Antebi}\ \emph {et~al.}(2021)\citenamefont {Antebi},
  \citenamefont {Pesin}, \citenamefont {Andreev},\ and\ \citenamefont
  {Ilan}}]{Antebi2021}%
  \BibitemOpen
  \bibfield  {author} {\bibinfo {author} {\bibfnamefont {O.}~\bibnamefont
  {Antebi}}, \bibinfo {author} {\bibfnamefont {D.~A.}\ \bibnamefont {Pesin}},
  \bibinfo {author} {\bibfnamefont {A.~V.}\ \bibnamefont {Andreev}},\ and\
  \bibinfo {author} {\bibfnamefont {R.}~\bibnamefont {Ilan}},\ }\bibfield
  {title} {\bibinfo {title} {{Anomaly-induced sound absorption in Weyl
  semimetals}},\ }\href {https://doi.org/10.1103/PhysRevB.103.214309}
  {\bibfield  {journal} {\bibinfo  {journal} {Phys. Rev. B}\ }\textbf {\bibinfo
  {volume} {103}},\ \bibinfo {pages} {214309} (\bibinfo {year}
  {2021})}\BibitemShut {NoStop}%
\bibitem [{\citenamefont {Sukhachov}\ and\ \citenamefont
  {Glazman}(2021)}]{Sukhachov2021}%
  \BibitemOpen
  \bibfield  {author} {\bibinfo {author} {\bibfnamefont {P.~O.}\ \bibnamefont
  {Sukhachov}}\ and\ \bibinfo {author} {\bibfnamefont {L.~I.}\ \bibnamefont
  {Glazman}},\ }\bibfield  {title} {\bibinfo {title} {{Anomalous sound
  attenuation in Weyl semimetals in magnetic and pseudomagnetic fields}},\
  }\href {https://doi.org/10.1103/PhysRevB.103.214310} {\bibfield  {journal}
  {\bibinfo  {journal} {Phys. Rev. B}\ }\textbf {\bibinfo {volume} {103}},\
  \bibinfo {pages} {214310} (\bibinfo {year} {2021})}\BibitemShut {NoStop}%
\bibitem [{\citenamefont {Furuseth}\ \emph {et~al.}(1973)\citenamefont
  {Furuseth}, \citenamefont {Brattås}, \citenamefont {Kjekshus}, \citenamefont
  {Enzell}, \citenamefont {Enzell},\ and\ \citenamefont
  {Swahn}}]{furuseth_crystal_1973}%
  \BibitemOpen
  \bibfield  {author} {\bibinfo {author} {\bibfnamefont {S.}~\bibnamefont
  {Furuseth}}, \bibinfo {author} {\bibfnamefont {L.}~\bibnamefont {Brattås}},
  \bibinfo {author} {\bibfnamefont {A.}~\bibnamefont {Kjekshus}}, \bibinfo
  {author} {\bibfnamefont {C.~R.}\ \bibnamefont {Enzell}}, \bibinfo {author}
  {\bibfnamefont {C.~R.}\ \bibnamefont {Enzell}},\ and\ \bibinfo {author}
  {\bibfnamefont {C.-G.}\ \bibnamefont {Swahn}},\ }\bibfield  {title} {\bibinfo
  {title} {{The {Crystal} {Structure} of {HfTe5}.}},\ }\href
  {https://doi.org/10.3891/acta.chem.scand.27-2367} {\bibfield  {journal}
  {\bibinfo  {journal} {Acta Chemica Scandinavica}\ }\textbf {\bibinfo {volume}
  {27}},\ \bibinfo {pages} {2367} (\bibinfo {year} {1973})}\BibitemShut
  {NoStop}%
\bibitem [{\citenamefont
  {Wang}(2021{\natexlab{a}})}]{chenjie_thermodynamically_2021}%
  \BibitemOpen
  \bibfield  {author} {\bibinfo {author} {\bibfnamefont {C.}~\bibnamefont
  {Wang}},\ }\bibfield  {title} {\bibinfo {title} {{Thermodynamically Induced
  Transport Anomaly in Dilute Metals ${\mathrm{ZrTe}}_{5}$ and
  ${\mathrm{HfTe}}_{5}$}},\ }\href
  {https://doi.org/10.1103/PhysRevLett.126.126601} {\bibfield  {journal}
  {\bibinfo  {journal} {Phys. Rev. Lett.}\ }\textbf {\bibinfo {volume} {126}},\
  \bibinfo {pages} {126601} (\bibinfo {year} {2021}{\natexlab{a}})}\BibitemShut
  {NoStop}%
\bibitem [{\citenamefont {Endo}\ \emph {et~al.}(2009)\citenamefont {Endo},
  \citenamefont {Hatano}, \citenamefont {Nakamura},\ and\ \citenamefont
  {Shirasaki}}]{Endo_2009}%
  \BibitemOpen
  \bibfield  {author} {\bibinfo {author} {\bibfnamefont {A.}~\bibnamefont
  {Endo}}, \bibinfo {author} {\bibfnamefont {N.}~\bibnamefont {Hatano}},
  \bibinfo {author} {\bibfnamefont {H.}~\bibnamefont {Nakamura}},\ and\
  \bibinfo {author} {\bibfnamefont {R.}~\bibnamefont {Shirasaki}},\ }\bibfield
  {title} {\bibinfo {title} {{Fundamental relation between longitudinal and
  transverse conductivities in the quantum Hall system}},\ }\href
  {https://doi.org/10.1088/0953-8984/21/34/345803} {\bibfield  {journal}
  {\bibinfo  {journal} {Journal of Physics: Condensed Matter}\ }\textbf
  {\bibinfo {volume} {21}},\ \bibinfo {pages} {345803} (\bibinfo {year}
  {2009})}\BibitemShut {NoStop}%
\bibitem [{\citenamefont {Klier}\ \emph {et~al.}(2015)\citenamefont {Klier},
  \citenamefont {Gornyi},\ and\ \citenamefont {Mirlin}}]{Klier_2015}%
  \BibitemOpen
  \bibfield  {author} {\bibinfo {author} {\bibfnamefont {J.}~\bibnamefont
  {Klier}}, \bibinfo {author} {\bibfnamefont {I.~V.}\ \bibnamefont {Gornyi}},\
  and\ \bibinfo {author} {\bibfnamefont {A.~D.}\ \bibnamefont {Mirlin}},\
  }\bibfield  {title} {\bibinfo {title} {{Transversal magnetoresistance in Weyl
  semimetals}},\ }\href {https://doi.org/10.1103/PhysRevB.92.205113} {\bibfield
   {journal} {\bibinfo  {journal} {Phys. Rev. B}\ }\textbf {\bibinfo {volume}
  {92}},\ \bibinfo {pages} {205113} (\bibinfo {year} {2015})}\BibitemShut
  {NoStop}%
\bibitem [{\citenamefont {K\"onye}\ and\ \citenamefont
  {Ogata}(2018)}]{koenye_magnetoresistance_2018}%
  \BibitemOpen
  \bibfield  {author} {\bibinfo {author} {\bibfnamefont {V.}~\bibnamefont
  {K\"onye}}\ and\ \bibinfo {author} {\bibfnamefont {M.}~\bibnamefont
  {Ogata}},\ }\bibfield  {title} {\bibinfo {title} {{Magnetoresistance of a
  three-dimensional Dirac gas}},\ }\href
  {https://doi.org/10.1103/PhysRevB.98.195420} {\bibfield  {journal} {\bibinfo
  {journal} {Phys. Rev. B}\ }\textbf {\bibinfo {volume} {98}},\ \bibinfo
  {pages} {195420} (\bibinfo {year} {2018})}\BibitemShut {NoStop}%
\bibitem [{\citenamefont {Hsu}\ and\ \citenamefont
  {Guo}(2010)}]{hsu_anomalous_2010}%
  \BibitemOpen
  \bibfield  {author} {\bibinfo {author} {\bibfnamefont {Y.-F.}\ \bibnamefont
  {Hsu}}\ and\ \bibinfo {author} {\bibfnamefont {G.-Y.}\ \bibnamefont {Guo}},\
  }\bibfield  {title} {\bibinfo {title} {{Anomalous integer quantum Hall effect
  in $\mathit{A}\mathit{A}$-stacked bilayer graphene}},\ }\href
  {https://doi.org/10.1103/PhysRevB.82.165404} {\bibfield  {journal} {\bibinfo
  {journal} {Phys. Rev. B}\ }\textbf {\bibinfo {volume} {82}},\ \bibinfo
  {pages} {165404} (\bibinfo {year} {2010})}\BibitemShut {NoStop}%
\bibitem [{\citenamefont {Ashby}\ and\ \citenamefont
  {Carbotte}(2014)}]{ashby_chiral_2014}%
  \BibitemOpen
  \bibfield  {author} {\bibinfo {author} {\bibfnamefont {P.~E.~C.}\
  \bibnamefont {Ashby}}\ and\ \bibinfo {author} {\bibfnamefont {J.~P.}\
  \bibnamefont {Carbotte}},\ }\bibfield  {title} {\bibinfo {title} {{Chiral
  anomaly and optical absorption in Weyl semimetals}},\ }\href
  {https://doi.org/10.1103/PhysRevB.89.245121} {\bibfield  {journal} {\bibinfo
  {journal} {Phys. Rev. B}\ }\textbf {\bibinfo {volume} {89}},\ \bibinfo
  {pages} {245121} (\bibinfo {year} {2014})}\BibitemShut {NoStop}%
\bibitem [{\citenamefont {Rhim}\ and\ \citenamefont
  {Kim}(2015)}]{rhim_quantum_2015}%
  \BibitemOpen
  \bibfield  {author} {\bibinfo {author} {\bibfnamefont {J.-W.}\ \bibnamefont
  {Rhim}}\ and\ \bibinfo {author} {\bibfnamefont {Y.~B.}\ \bibnamefont {Kim}},\
  }\bibfield  {title} {\bibinfo {title} {{Quantum oscillations in the Luttinger
  model with quadratic band touching: Applications to pyrochlore iridates}},\
  }\href {https://doi.org/10.1103/PhysRevB.91.115124} {\bibfield  {journal}
  {\bibinfo  {journal} {Phys. Rev. B}\ }\textbf {\bibinfo {volume} {91}},\
  \bibinfo {pages} {115124} (\bibinfo {year} {2015})}\BibitemShut {NoStop}%
\bibitem [{\citenamefont {Bruus}\ and\ \citenamefont
  {Flensberg}(2004)}]{bruus_manybody_2004}%
  \BibitemOpen
  \bibfield  {author} {\bibinfo {author} {\bibfnamefont {H.}~\bibnamefont
  {Bruus}}\ and\ \bibinfo {author} {\bibfnamefont {K.}~\bibnamefont
  {Flensberg}},\ }\href@noop {} {\emph {\bibinfo {title} {{Many-body quantum
  theory in condensed matter physics - an introduction}}}}\ (\bibinfo
  {publisher} {Oxford University Press},\ \bibinfo {address} {United States},\
  \bibinfo {year} {2004})\BibitemShut {NoStop}%
\bibitem [{\citenamefont {Zhang}\ \emph
  {et~al.}(2019{\natexlab{b}})\citenamefont {Zhang}, \citenamefont {Song},
  \citenamefont {Nie}, \citenamefont {Liu}, \citenamefont {Wang}, \citenamefont
  {Jiang}, \citenamefont {Zhao}, \citenamefont {Chen}, \citenamefont {Meng},
  \citenamefont {Duan},\ and\ \citenamefont {Liu}}]{zhang_ultrafast_2019}%
  \BibitemOpen
  \bibfield  {author} {\bibinfo {author} {\bibfnamefont {X.}~\bibnamefont
  {Zhang}}, \bibinfo {author} {\bibfnamefont {H.-Y.}\ \bibnamefont {Song}},
  \bibinfo {author} {\bibfnamefont {X.~C.}\ \bibnamefont {Nie}}, \bibinfo
  {author} {\bibfnamefont {S.-B.}\ \bibnamefont {Liu}}, \bibinfo {author}
  {\bibfnamefont {Y.}~\bibnamefont {Wang}}, \bibinfo {author} {\bibfnamefont
  {C.-Y.}\ \bibnamefont {Jiang}}, \bibinfo {author} {\bibfnamefont {S.-Z.}\
  \bibnamefont {Zhao}}, \bibinfo {author} {\bibfnamefont {G.}~\bibnamefont
  {Chen}}, \bibinfo {author} {\bibfnamefont {J.-Q.}\ \bibnamefont {Meng}},
  \bibinfo {author} {\bibfnamefont {Y.-X.}\ \bibnamefont {Duan}},\ and\
  \bibinfo {author} {\bibfnamefont {H.~Y.}\ \bibnamefont {Liu}},\ }\bibfield
  {title} {\bibinfo {title} {{Ultrafast hot carrier dynamics of
  ${\mathrm{ZrTe}}_{5}$ from time-resolved optical reflectivity}},\ }\href
  {https://doi.org/10.1103/PhysRevB.99.125141} {\bibfield  {journal} {\bibinfo
  {journal} {Phys. Rev. B}\ }\textbf {\bibinfo {volume} {99}},\ \bibinfo
  {pages} {125141} (\bibinfo {year} {2019}{\natexlab{b}})}\BibitemShut
  {NoStop}%
\bibitem [{\citenamefont {Wawrzy{\'n}czak}\ \emph {et~al.}(2021)\citenamefont
  {Wawrzy{\'n}czak}, \citenamefont {Galeski}, \citenamefont {Noky},
  \citenamefont {Sun}, \citenamefont {Felser},\ and\ \citenamefont
  {Gooth}}]{wawrzynczak_three_2021}%
  \BibitemOpen
  \bibfield  {author} {\bibinfo {author} {\bibfnamefont {R.}~\bibnamefont
  {Wawrzy{\'n}czak}}, \bibinfo {author} {\bibfnamefont {S.}~\bibnamefont
  {Galeski}}, \bibinfo {author} {\bibfnamefont {J.}~\bibnamefont {Noky}},
  \bibinfo {author} {\bibfnamefont {Y.}~\bibnamefont {Sun}}, \bibinfo {author}
  {\bibfnamefont {C.}~\bibnamefont {Felser}},\ and\ \bibinfo {author}
  {\bibfnamefont {J.}~\bibnamefont {Gooth}},\ }\bibfield  {title} {\bibinfo
  {title} {{Three-dimensional quasi-quantized Hall effect in bulk InAs}},\
  }\href@noop {} {\bibfield  {journal} {\bibinfo  {journal} {arXiv preprint
  arXiv:2102.04928}\ } (\bibinfo {year} {2021})}\BibitemShut {NoStop}%
\bibitem [{\citenamefont {Xiao}\ \emph {et~al.}(2017)\citenamefont {Xiao},
  \citenamefont {Law},\ and\ \citenamefont
  {Lee}}]{xiao_magnetoconductivity_2017}%
  \BibitemOpen
  \bibfield  {author} {\bibinfo {author} {\bibfnamefont {X.}~\bibnamefont
  {Xiao}}, \bibinfo {author} {\bibfnamefont {K.~T.}\ \bibnamefont {Law}},\ and\
  \bibinfo {author} {\bibfnamefont {P.~A.}\ \bibnamefont {Lee}},\ }\bibfield
  {title} {\bibinfo {title} {{Magnetoconductivity in Weyl semimetals: Effect of
  chemical potential and temperature}},\ }\href
  {https://doi.org/10.1103/PhysRevB.96.165101} {\bibfield  {journal} {\bibinfo
  {journal} {Phys. Rev. B}\ }\textbf {\bibinfo {volume} {96}},\ \bibinfo
  {pages} {165101} (\bibinfo {year} {2017})}\BibitemShut {NoStop}%
\bibitem [{\citenamefont {Klier}\ \emph {et~al.}(2017)\citenamefont {Klier},
  \citenamefont {Gornyi},\ and\ \citenamefont
  {Mirlin}}]{klier_transversal_2017}%
  \BibitemOpen
  \bibfield  {author} {\bibinfo {author} {\bibfnamefont {J.}~\bibnamefont
  {Klier}}, \bibinfo {author} {\bibfnamefont {I.~V.}\ \bibnamefont {Gornyi}},\
  and\ \bibinfo {author} {\bibfnamefont {A.~D.}\ \bibnamefont {Mirlin}},\
  }\bibfield  {title} {\bibinfo {title} {{Transversal magnetoresistance and
  Shubnikov--de Haas oscillations in Weyl semimetals}},\ }\href
  {https://doi.org/10.1103/PhysRevB.96.214209} {\bibfield  {journal} {\bibinfo
  {journal} {Phys. Rev. B}\ }\textbf {\bibinfo {volume} {96}},\ \bibinfo
  {pages} {214209} (\bibinfo {year} {2017})}\BibitemShut {NoStop}%
\bibitem [{\citenamefont
  {Wang}(2021{\natexlab{b}})}]{wang_thermodynamically_2021}%
  \BibitemOpen
  \bibfield  {author} {\bibinfo {author} {\bibfnamefont {C.}~\bibnamefont
  {Wang}},\ }\bibfield  {title} {\bibinfo {title} {{Thermodynamically Induced
  Transport Anomaly in Dilute Metals ${\mathrm{ZrTe}}_{5}$ and
  ${\mathrm{HfTe}}_{5}$}},\ }\href
  {https://doi.org/10.1103/PhysRevLett.126.126601} {\bibfield  {journal}
  {\bibinfo  {journal} {Phys. Rev. Lett.}\ }\textbf {\bibinfo {volume} {126}},\
  \bibinfo {pages} {126601} (\bibinfo {year} {2021}{\natexlab{b}})}\BibitemShut
  {NoStop}%
\end{thebibliography}%

\end{document}